\documentclass{article}

\usepackage{arxiv}

\usepackage[utf8]{inputenc} 
\usepackage[T1]{fontenc}    
\usepackage{url}            
\usepackage{booktabs}       
\usepackage{amsfonts,amsmath,amsthm}       
\usepackage{nicefrac}       
\usepackage{graphicx}
\usepackage{subfigure}
\graphicspath{ {./images/} }

\usepackage{color}
\usepackage[english]{babel}
\usepackage{amsmath,amssymb,amsthm}
\usepackage{subfigure}
\usepackage{multirow}
\usepackage{ctable}
\usepackage[normalem]{ulem}

\newcommand{\B}[1]{\mbox{\boldmath${#1}$\unboldmath}} 
\def\R{\mathbb{R}}

\newcommand{\sgn}{\text{sgn}}

\title{Data-driven intrinsic localized mode detection and classification in one-dimensional crystal lattice model\thanks{Preprint.}}

\author{
 J\={a}nis Baj\={a}rs \\
 Faculty of Physics, Mathematics and Optometry \\
 University of Latvia \\
 Jelgavas Street 3\\
 Riga, LV-1004, Latvia \\
 \texttt{janis.bajars@lu.lv}  
 \And
 Filips Kozirevs \\
 Faculty of Physics, Mathematics and Optometry \\
 University of Latvia \\
 Jelgavas Street 3\\
 Riga, LV-1004, Latvia \\
 \texttt{filips.kozirevs@lu.lv}
}

\begin{document}

\maketitle

\begin{abstract}
In this work we propose Support Vector Machine classification algorithms to classify one-dimensional crystal lattice waves from locally sampled data. Different learning datasets of particle displacements, momenta and energy density values are considered. Efficiency of the classification algorithms is further improved by two dimensionality reduction techniques: Principal Component Analysis and Locally Linear Embedding. Robustness of classifiers is investigated and demonstrated. Developed algorithms are successfully applied to detect localized intrinsic modes in three numerical simulations considering a case of two localized stationary breather solutions, a single stationary breather solution in noisy background and two mobile breather collision.   
\end{abstract}


\section{Introduction}\label{sec:Intro}

With constant increase of computational power data-driven methods have gained a significant popularity in the past few decades with applications spanning all fields of science \cite{Montans19}. The study of {\it intrinsic localized modes} (ILMs) in crystal lattice models, also known as {\it discrete breathers} (DBs), is an active field of research in mathematics and physics \cite{MacKay94,Flach08,ProcMica14,ram20} which relies heavily on obtained data from numerical simulations. Thus, emerging data-driven methods provide great opportunity to explore new techniques for understanding nonlinear wave phenomena in crystal lattice models.  

Such localized wave solutions are of particular interest in material science where transport of charge in silicates by moving nonlinear localized modes is experimentally confirmed, the phenomenon called {\it hyperconductivity} \cite{ram20,Russell19}. In addition, localized modes provide mechanisms for the formation of long decorated dark lines in muscovite mica \cite{ProcMica14} and can have a functional role in proteins \cite{Senet15,Piazza08}. Strong evidence of localized phenomena in crystal lattice models is provided by numerical experiments \cite{Chris11,Marin01,Bajars15,Shepelev20,Chetverikov16,Bajars21}, to mention but a few.

There are already well established theoretical and numerical understandings of the existence of one-dimensional stationary DBs \cite{MacKay94,Flach08,Aubry06}, i.e., spatially localized time-periodic excitations. The same may not be said about propagating DBs \cite{Flach99} and, in particular, in higher dimensions. There are still open theoretical questions regarding the existence and properties of propagating DBs in general nonintegrable lattices. Mobile discrete breathers are not expected to be long-lived, since the lattice models considered are likely to be nonintegrable and their lifespan is subject to interactions with lattice defects and the phonon background, which may be viewed as thermal noise. Thus, an obvious challenge is to study the existence and interactions of long-lived propagating discrete breather solutions \cite{Bajars15,Bajars21} to understand their role in the transport of energy in sputtering experiments \cite{Russell07}. Such demanding computations require good data-based numerical tools to detect and track breather solutions in space and time which serves as a motivation for our work. 

In this work we develop classification algorithms to differentiate between nonlinear localized waves and nonlocalized linear waves in numerical simulations of one-dimensional crystal lattice model. Obtained classifiers rely only on locally sampled data at a given time in opposed to nonlocal time series analysis methods, such as discrete Fourier and wavelet transforms. For instance, discrete Fourier transform in space and time provides frequency-momentum representation of the lattice waves and allows characterization of exact ILMs \cite{Juan19,JBJA22} but eliminates all space-time domain information and does not provide any means to locate the localization regions in space at a given time. On the other hand, localization of frequency and time of a single particle dynamics can be studied with the G\'{a}bor transformation, also known as short-time Fourier transform, with localization accuracy subject to the Heisenberg uncertainty relationship, i.e., there are trade-offs between time and frequency resolutions. Wavelet transforms, which are extensions of the G\'{a}bor transformation, subsequently shorten the scaling window to extract
higher frequencies at better time resolution. Wavelet analysis of ILMs in the time-frequency domain has been studied in \cite{Forinash98,Annise19}, while authors in \cite{Hori93} considered wavelet analysis in the space-wavenumber domain. Fourier, G\'{a}bor and wavelet transforms provide valuable spectral information and its evolution in time or space. To determine the actual localization sites at a given time in this work we propose sliding window object detection method together with developed classifiers.

Such classifiers can be efficiently trained with different given labeled datasets which are not only limited to numerical simulation data. Trained classifiers then can be used to detect localization regions, e.g., in numerical simulations, which is the first step towards fully automated tool for quantitative data-based analysis of complex numerical experiments. Importantly, our analysis and methodology, which follows, extends, in general, to any one-dimensional crystal lattice models which support intrinsic localized mode solutions.    

The paper is organized in the following way. In Section \ref{sec:MathModel} mathematical model of one-dimensional crystal lattice is formulated. Spectral properties of discrete lattice waves are discussed in Section \ref{sec:Spectrum}. Dimensionality reduction methods and classification algorithms are presented in Section \ref{sec:Classification}. In Section \ref{sec:Applications} application of the classification algorithms is demonstrated for detecting localized modes in numerical simulations. Discussion and conclusions are provided in Section \ref{sec:Conclusions}.    

\section{Mathematical model}\label{sec:MathModel}

In this work we are concerned with one-dimensional crystal lattice models of classical molecular dynamics exhibiting ILM solutions. In particular, we consider Hamiltonian dynamics of $N$ particles arising from the dimensionless Hamiltonian:
\begin{equation}\label{eq:Hamilt}
H = K + U + V 
  = \sum_{n=1}^{N} \left( \frac{1}{2} \dot{q}_{n}^2 +
  U(q_{n}) + V(|q_{n+1}-q_{n}|) \right),
\end{equation}
where $K$ is the kinetic energy, $U$ is the on-site potential energy, $V$ is the interaction potential of particles, $q_n\in\mathbb{R}$ is the position of the $n^{th}$ particle, $\dot{q}_{n}$ is its time derivative and $|\cdot|$ is the Euclidean distance. In the dimensionless form \eqref{eq:Hamilt} particle equilibrium distance $\Delta=q_{n+1}-q_n$ as well as all masses are equal to one. 

For example, such model \eqref{eq:Hamilt} arises as one-dimensional model of the K-K layer of layered silicate muscovite mica \cite{Chris11} or simplified model of two-dimensional hexagonal crystal lattice \cite{Bajars15,Bajars21}. In such case, the on-site potential $U$ models forces from atoms above and below the K-K sheet. Thus, the on-site potential is modeled as smoothed periodic function \cite{Chris11}:
\begin{equation}\label{eq:OnSite}
U(q_n) = 1-\cos(2 \pi q_n).
\end{equation}
Note that more realistic potentials for muscovite mica can also be considered \cite{Juan19}.

There are multiple options for modeling the interaction potential $V$ \cite{Allen89}. In this work we consider well known scaled Lennard-Jones potential in the following form:
\begin{equation}\label{eq:LJ}
V(r_n) = \epsilon \left( \left( \frac{1}{r_n} \right)^{12} - 2 \left(
    \frac{1}{r_n} \right)^{6} \right), \quad r_n = |q_{n+1}-q_n|,
\end{equation}
where the parameter $\epsilon>0$ describes the ratio of the atomic well depths associated to the interaction and the on-site potential. That is, large values of $\epsilon$ lead to a model describing one-dimensional Lennard-Jones fluid, while small values of $\epsilon$ lead to system of decoupled nonlinear oscillators. Further simplifications in \eqref{eq:Hamilt} have been made to only consider close neighbor interactions with periodic boundary conditions, i.e., ${q}_{N+1} = {q}_{1}$. Note that proposed methodology of Section \ref{sec:Classification} is not limited to the particular choice of potentials \eqref{eq:OnSite}--\eqref{eq:LJ} and only fixed neighbor interactions.   

From the Hamiltonian \eqref{eq:Hamilt} obtained Hamiltonian equations:
\begin{align}
\dot{q}_n & = p_n, \label{eq:q}\\
\dot{p}_n & = -U'(q_n) + V'(|q_{n+1}-q_{n}|) - V'(|q_{n}-q_{n-1}|), \label{eq:p}
\end{align}
where $p_n$ is momentum, are solved numerically with the second order time reversible symplectic Verlet method \cite{Hairer}. In the following, without loss of generality, all analysis and numerical simulations are performed with time step $\tau=0.01$ and $\epsilon=0.05$. 

To excite DB solutions we consider particles in their dynamical equilibrium states, i.e., $q^0_n=n-1$ and $p^0_n=0$, while exciting four neighboring particle momenta with the pattern:
\begin{equation}\label{eq:patern}
\B{p}_0 = \gamma \, (-1, 2, -2, 1)^T, \quad \gamma > 0.
\end{equation} 
Such excitation, due to nonlinearity, generates in our model stationary discrete breather. In the current study we only consider numerical data from stationary DB solutions while the extension to differentiate between stationary and moving breather solutions is left for future work.

In general, larger values of $\gamma$ generate localized waves with larger particle displacements. While such excitation \eqref{eq:patern} does not generate exact wave solutions, the amount of generated linear phonon waves has been observed to be minimal. Not all values of $\gamma$ can be considered. Too small values will generate phonons, while too large values will force particles to move out from their potential wells or in addition to the stationary breather will also generate moving breathers. Moreover, appropriate values of $\gamma$ depend on $\epsilon$. With spectral analysis of Section \ref{sec:Spectrum} and $\epsilon=0.05$ we find $\gamma$ lower bound $0.15$ for the initial momenta pattern \eqref{eq:patern} to generate stationary DB. On the other hand, to generate arbitrary phonon waves we can choose small random momenta and particle displacement values as initial conditions, as confirmed by the spectral analysis in Section \ref{sec:Spectrum}. 

\begin{figure}[t]
\centering 
\subfigure[]{\label{fig:Bsol}
\includegraphics[trim=1cm 0.1cm 1cm 1cm,clip=true,width=0.48\textwidth]{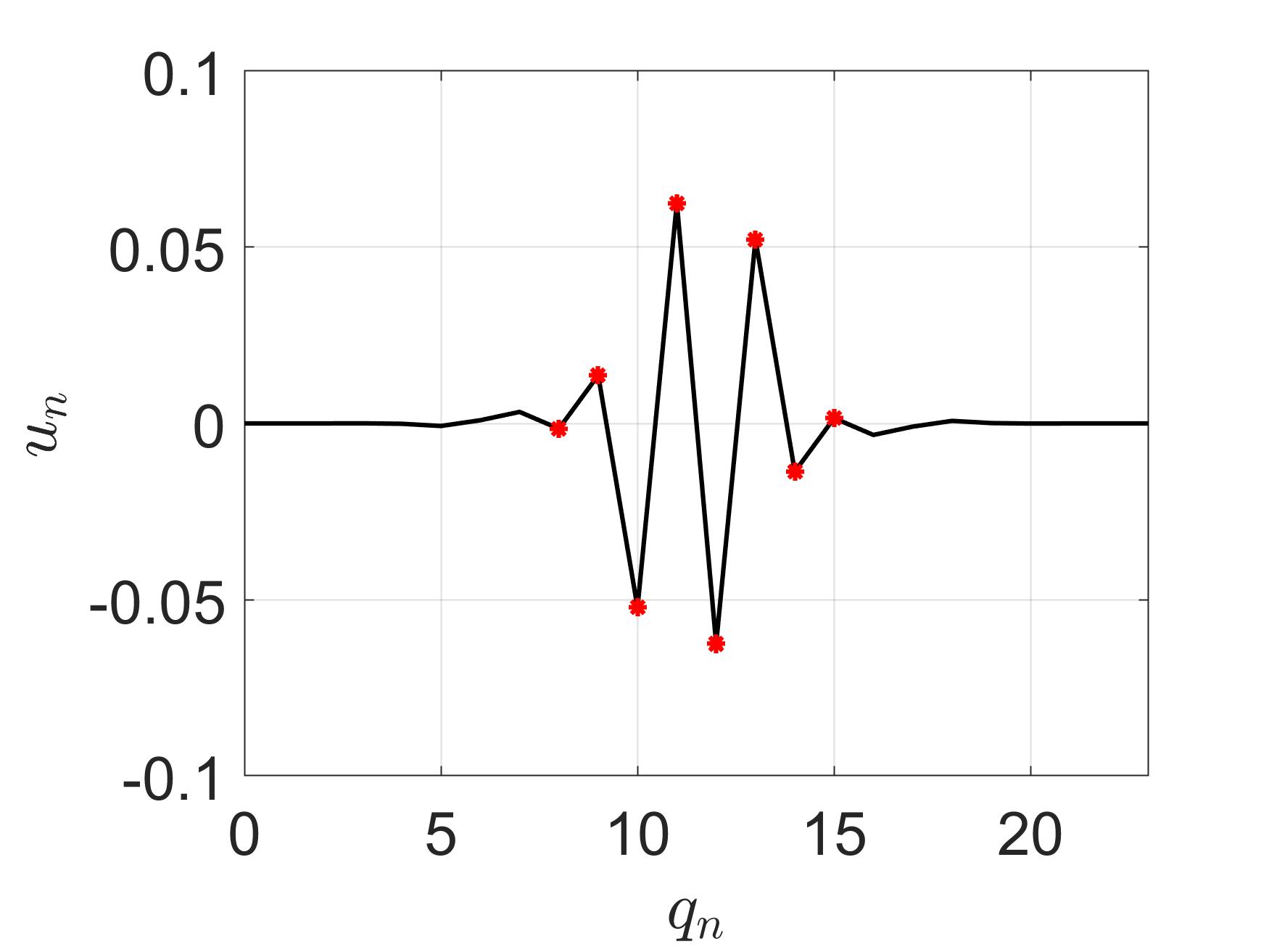}}
\subfigure[]{\label{fig:Psol}
\includegraphics[trim=1cm 0.1cm 1cm 1cm,clip=true,width=0.48\textwidth]{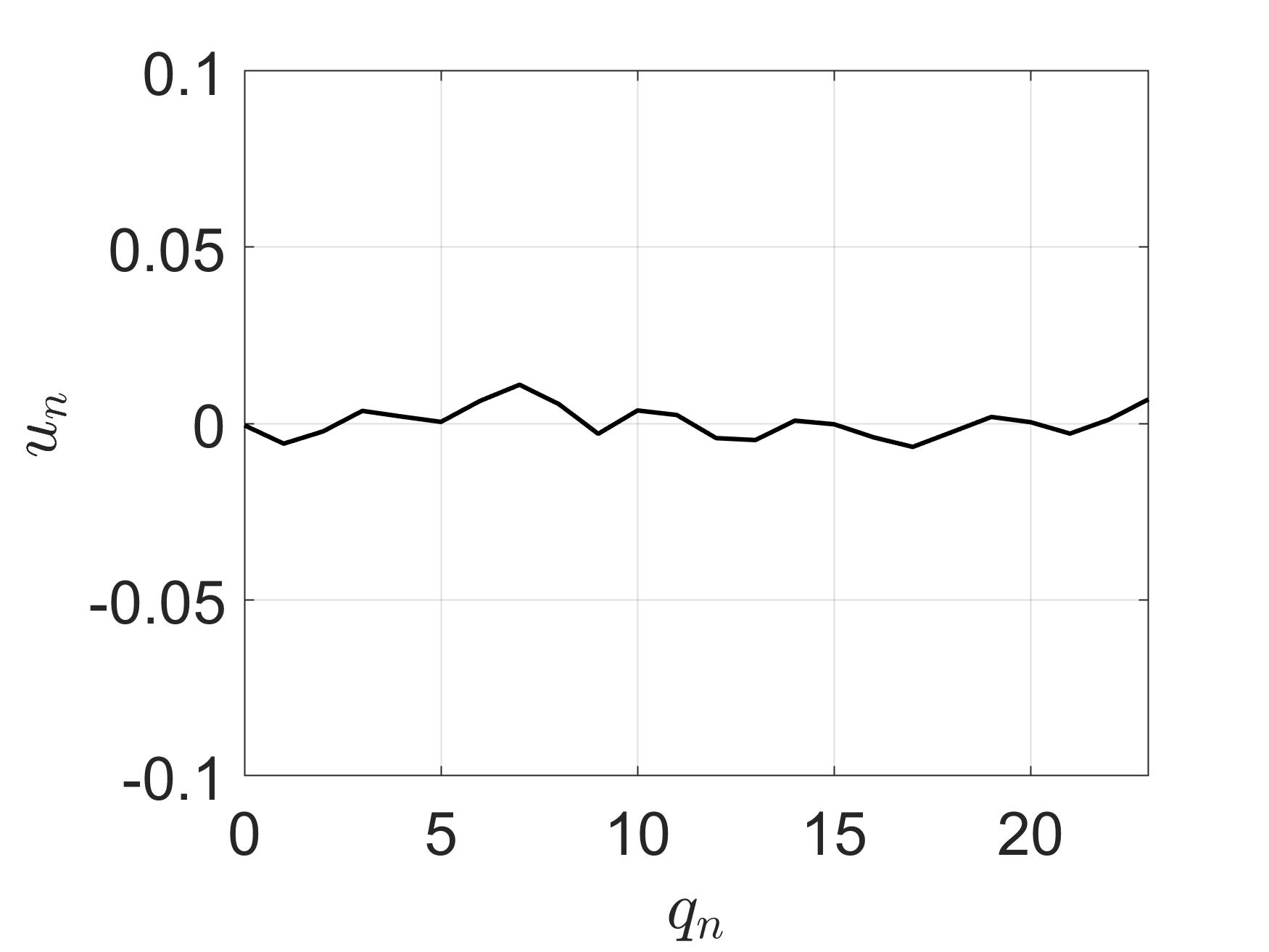}}
\caption{Particle displacements from the numerical simulations at $T_{end}=5$ with $\epsilon=0.05$ and $N=24$. (a) stationary discrete breather solution initiated with $\gamma=0.3$ and markers indicating eight particles on which the breather is localized. (b) phonon wave solution which does not exhibit localization.}\label{fig:sol}
\end{figure}

Both described initial conditions will be used to generate datasets in Section \ref{sec:Classification} to obtain classification and detection algorithms which are capable of distinguishing between localized and phonon waves. To see discrepancies between both types of waves in Figure \ref{fig:sol} we illustrate particle displacement function $u_n = q_n - q_n^0$ from two numerical simulations with $N=24$ at final computational time $T_{end}=5$. In Figure \ref{fig:Bsol} we show displacement function of a localized discrete breather solution computed with $\gamma=0.3$, while in Figure \ref{fig:Psol} we demonstrate particle displacements for phonon waves computed with random initial momentum and particle displacement values. It is easy to see larger displacement values for the breather solution and that it is spatially localized compared to the approximately linear waves. In addition, in Figure \ref{fig:Bsol} we have added markers to eight particle displacement values on which the breather is localized. Exactly eight data point values will be used in the classification algorithms of Section \ref{sec:Classification}.

\section{Spectral properties of lattice waves} \label{sec:Spectrum}

It is well known that linear phonon waves in crystal lattice models can be characterized by their dispersion relation \cite{Flach08}. For the model \eqref{eq:q}--\eqref{eq:p} linearization at the equilibrium $(q_n^0,p_n^0)=(n-1,0)$ yields the linearized equations:
\begin{align}\label{eq:LinEq}
\ddot{u}_n & = -U''(q_n^0)u_n + V''(r_n^0)u_{n+1} - V''(r_n^0)u_{n} 
- V''(r_{n-1}^0)u_{n} + V''(r_{n-1}^0)u_{n-1} \nonumber \\
 & = -U''(q_n^0)u_n + V''(1)(u_{n+1} - 2u_{n} + u_{n-1}) \nonumber \\
 & = - \omega_0^2 u_n + c^2 (u_{n+1} - 2u_{n} + u_{n-1}), 
\end{align}
where $r_{n}^0=|q_{n+1}^0-q_n^0|=1$, $r_{n-1}^0=|q_{n}^0-q_{n-1}^0|=1$, $\omega_0^2:=U''(q_n^0)=4\pi^2$ and $c^2:=V''(1)=72\epsilon$. With ansatz of plane wave solutions $u_n = \exp(\mathrm{i}(kn-\omega t))$ we derive the dispersion relation:
\begin{equation}\label{eq:dispersion}
\omega^2 = \omega_0^2 + 4 c^2 \sin^2 \left( \frac{k}{2} \right),
\end{equation}
where $\omega$ and $k$ are the phonon frequency and wave number, respectively, $\omega_0$ is the oscillation frequency of the isolated oscillators when $\epsilon=0$ and $c$ is the speed of sound when $\omega_0=0$ and $k\to 0$ which coincides with the maximal phase and group velocity values, i.e., $V_{ph}=c\frac{\sin(k/2)}{k/2}$ and $V_{g}=c\cos(k/2)$, respectively. In the case with the on-site potential $(\omega_0 \neq 0)$, according to \cite{Juan19}, since $V_{ph}\to\infty$ when $k\to 0$, the more appropriate definition of the sound speed is the maximum of the group velocity $V_g=c^2\sin(k)/\omega$ which is $V_g \approx 0.5285$ at $k \approx 0.4753 \pi$ if $\epsilon=0.05$.

\begin{figure}[t]
\centering 
\subfigure[]{\label{fig:SPb}
\includegraphics[trim=0.4cm 0.5cm 0.6cm 1cm,clip=true,width=0.48\textwidth]{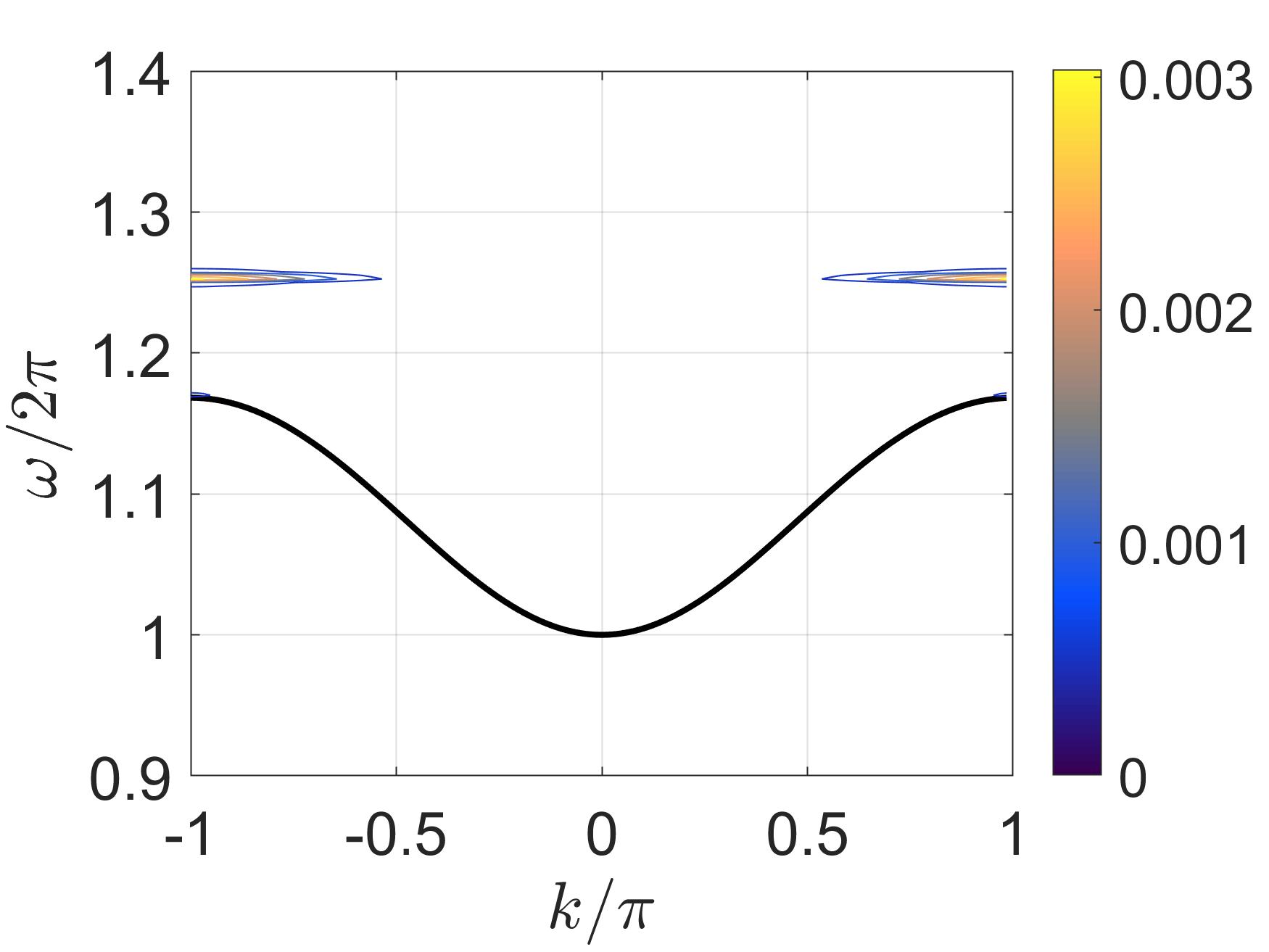}}
\subfigure[]{\label{fig:SPp}
\includegraphics[trim=0.4cm 0.5cm 0.6cm 1cm,clip=true,width=0.48\textwidth]{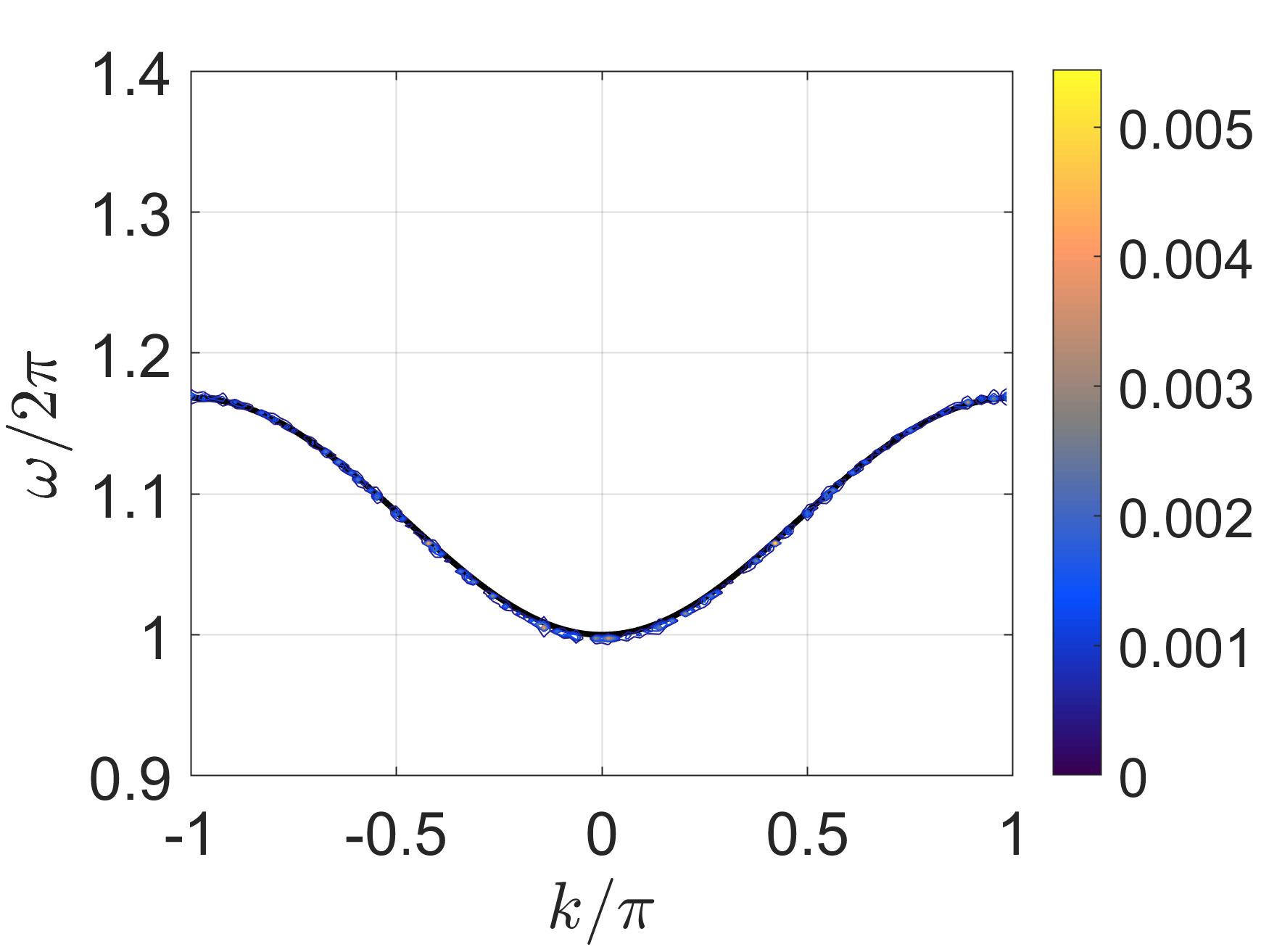}}
\caption{Contours of two-dimensional discrete Fourier transform in space and time of numerical simulation with $N=128$ and $T_{end}=400$. Dispersion relation \eqref{eq:dispersion} is indicated by the solid line. (a) stationary discrete breather amplitude spectrum computed with the initial pattern \eqref{eq:patern} and $\gamma=0.3$. (b) phonon wave amplitude spectrum computed with random initial condition and coinciding with the dispersion curve \eqref{eq:dispersion}.}\label{fig:SP}
\end{figure}

The dispersion relation \eqref{eq:dispersion} is optic-like \cite{Flach08} since it is bounded from below by $\omega_0=2\pi$ at $k=0$. In addition, it is also bounded from above at $k=\pm \pi$ as can be seen in Figure \ref{fig:SP}. In Figure \ref{fig:SP} we plot two-dimensional amplitude spectrum of numerical simulation data in space and time, i.e., contours of 2D discrete Fourier transform, where the color axis indicate Fourier mode amplitude values. We consider a lattice of $N=128$ particles and integrate in time until $T_{end}=400$. The solid line indicates the dispersion relation \eqref{eq:dispersion}. 

Excited stationary breather solutions by \eqref{eq:patern} are high-frequency optical breathers, meaning that their frequencies are above the optic-like phonon linear spectrum as can be seen in Figure \ref{fig:SPb}, with adjacent particles moving out of phase. Only small amount of energy is localized on the dispersion curve indicating minimal presence of phonons in numerical simulation. It is easy to see that the stationary DB is not only localized in space, see Figure \ref{fig:Bsol}, but also in frequency, a so called {\it fundamental frequency} of an exact time periodic localized in space wave solution \cite{Juan19}, which in our model varies depending on the value of $\gamma$.

In Figure \ref{fig:SPp} we considered random initial conditions for the momentum and displacement variables $p_n$ and $u_n$, respectively, drawn from the uniform distribution $\mathcal{U}(-0.01,0.01)$. Obtained momentum values are then rescaled such that the total energy values \eqref{eq:Hamilt} are the same in both simulations. Such random initial condition leads to full spectrum of phonons which coincides with the dispersion curve \eqref{eq:dispersion}. Thus, this provides representative data of phonons for data analysis. 

Spectral analysis of this section not only provides insights into the lattice wave properties but also allows to justify the methods for obtaining so called labeled training and testing data for classification problem presented in the following section.

\section{Classification of lattice waves} \label{sec:Classification}

In this section we describe classification algorithms of crystal lattice waves based on locally sampled data. Machine learning techniques \cite{ML,Bishop} are considered for data dimensionality reduction and building an efficient classifier. All calculations are performed in {\it Python} with built-in functions of the open source {\it scikit-learn}\footnote{\url{https://scikit-learn.org/stable/}} library.

\begin{figure}[t]
\centering 
\subfigure[]{\label{fig:Enb}
\includegraphics[trim=0.5cm 0.4cm 0.8cm 1cm,clip=true,width=0.48\textwidth]{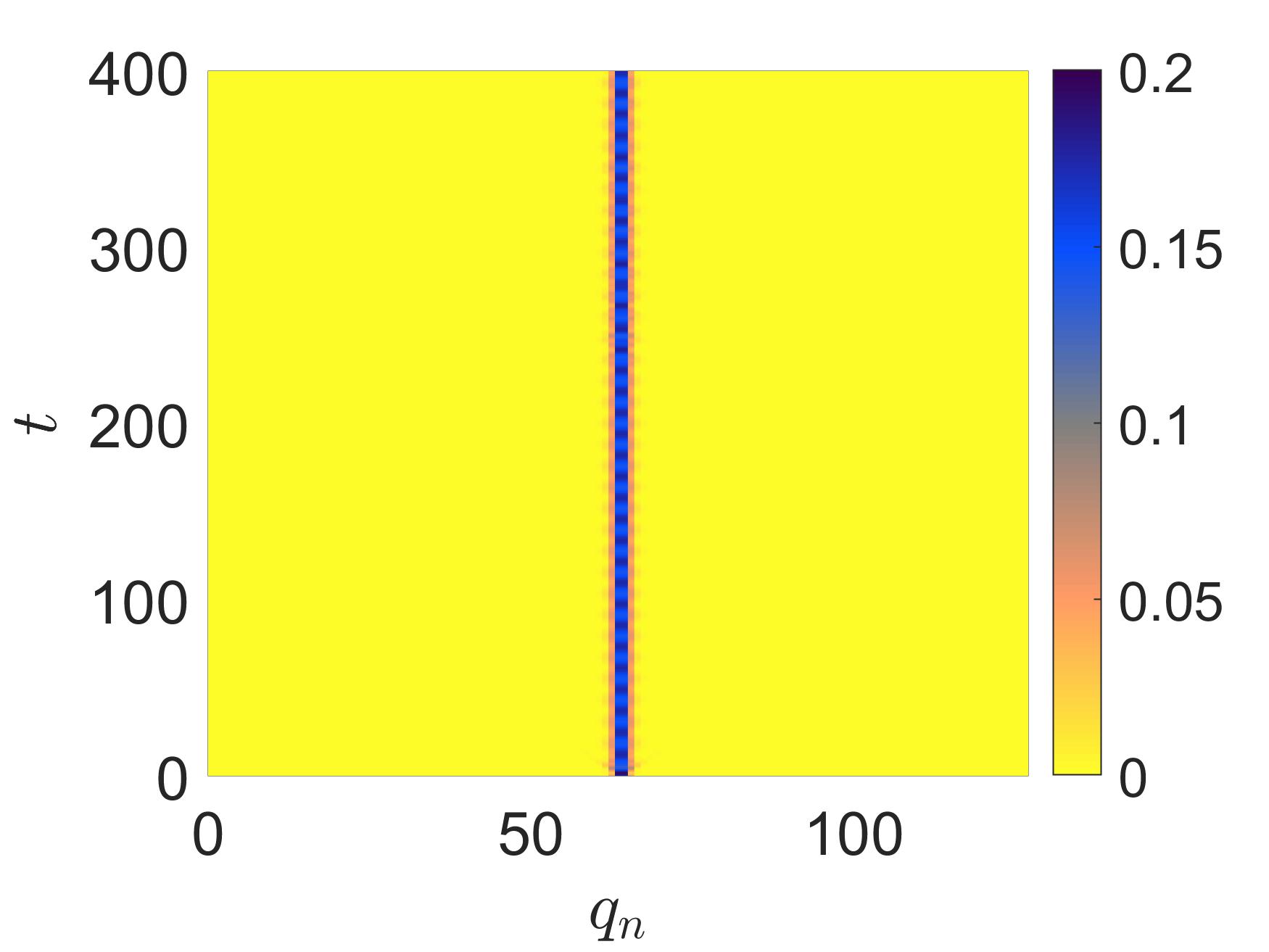}}
\subfigure[]{\label{fig:Enp}
\includegraphics[trim=0.5cm 0.4cm 0.8cm 1cm,clip=true,width=0.48\textwidth]{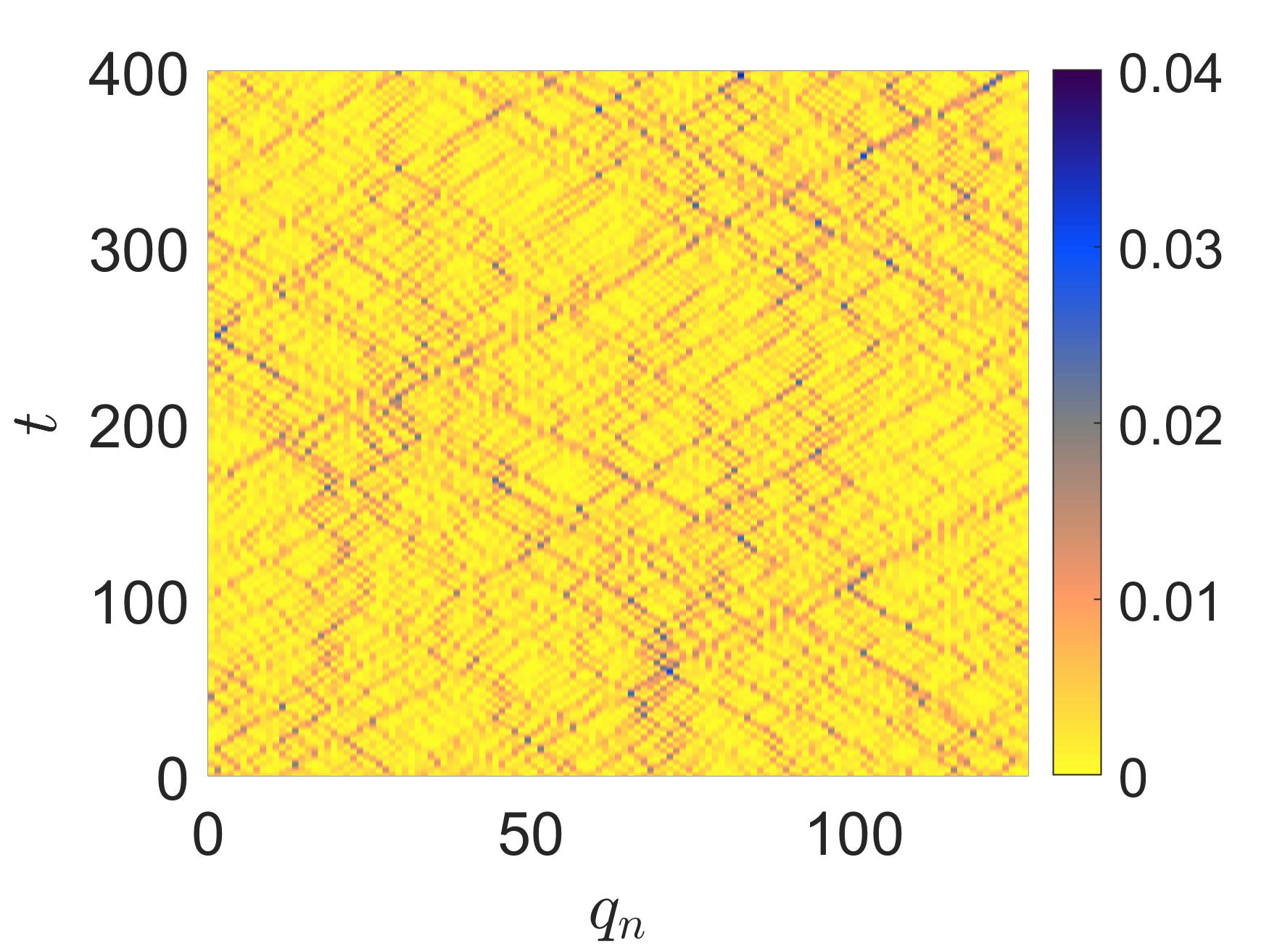}}
\caption{Energy density $E_n+\epsilon$ in time of two numerical simulations of Figure \ref{fig:SP} with $N=128$ and $T_{end}=400$. (a) stationary discrete breather simulation with the initial pattern \eqref{eq:patern} and $\gamma=0.3$. (b) phonon wave simulation with random initial condition.}\label{fig:En}
\end{figure}

We differentiate localized discrete breather solutions, see Figure \ref{fig:Bsol}, from nonlocalized linear phonons, see Figure \ref{fig:Psol}. Training (including validation) and testing data is obtained from total of $N_{sim}=1000$ numerical simulations with $0.75$ localized and nonlocalized wave data proportion. Local data of particle displacements, momenta and energies is sampled from eight neighboring particles indicated by red dots in Figure \ref{fig:Bsol}. For presentation purposes we consider three different datasets for classification: $X_{u}$, $X_{u,p}$ and $X_{u,p,E}$, where the dataset $X_{u}$ only contains particle displacement $u_n$ values, the dataset $X_{u,p}$ contains particle displacement $u_n$ and momenta $p_n$ values, while the dataset $X_{u,p,E}$ in addition to displacement and momenta values contains also particle energy density data defined by
\begin{equation}\label{eq:En}
E_n := \frac{1}{2} {p}_{n}^2 + U(q_{n}) + \frac{1}{2}\left( V(|q_{n}-q_{n-1}|) + V(|q_{n+1}-q_{n}|) \right).
\end{equation}
Considering available numerical data, $u_n$, $p_n$ and $E_n$, we have explored all different combinations, i.e., in total seven different datasets. Not all results are shown but the main differences are stated. Since we collect data from eight neighboring particles, i.e., $N_d=8$, we have that
\begin{equation}\label{eq:datasets}
X_{u} \in \R^{N_{sim} \times N_d}, \quad 
X_{u,p} \in \R^{N_{sim} \times 2N_d}, \quad 
X_{u,p,E} \in \R^{N_{sim} \times 3N_d}.
\end{equation}
For further reference we define $K:=jN_d$, where $j=1$ or $j=2$, or $j=3$. Thus, any given dataset is a matrix $X\in \R^{N_{sim}\times K}$.

To appreciate the importance of the particle energy density values \eqref{eq:En}, which also will be evident in the following sections, in Figure \ref{fig:En} we illustrate particle energy density function of two numerical simulations of Figure \ref{fig:SP}. In the illustration we have added constant $\epsilon$ to $E_n$ such that the minimal value is equal to zero. It is easy to see that the localized wave in Figure \ref{fig:Enb} carries higher value of energy in the localized region compared to nonlocalized phonon waves in Figure \ref{fig:Enp}. Thus, energy value is an important quantity to consider in datasets for building a classifier. 

In what follows, we excite discrete breather solutions with momenta pattern \eqref{eq:patern} where parameter $\gamma$ values are randomly chosen from the uniform distribution $\mathcal{U}(0.15,1.2)$. Linear phonon waves are generated with random initial particle displacement and momenta values from the uniform distribution $\mathcal{U}(-0.01,0.01)$. Data is collected after performing numerical integration of \eqref{eq:q}--\eqref{eq:p} in time until $T_{end}=3$ and in lattice with $N=64$ particles.   

\subsection{Dimensionality reduction}

\begin{figure}[t]
\centering 
\subfigure[]{\label{fig:corrMb}
\includegraphics[trim=0cm 0cm 1.1cm 1cm,clip=true,width=0.48\textwidth]{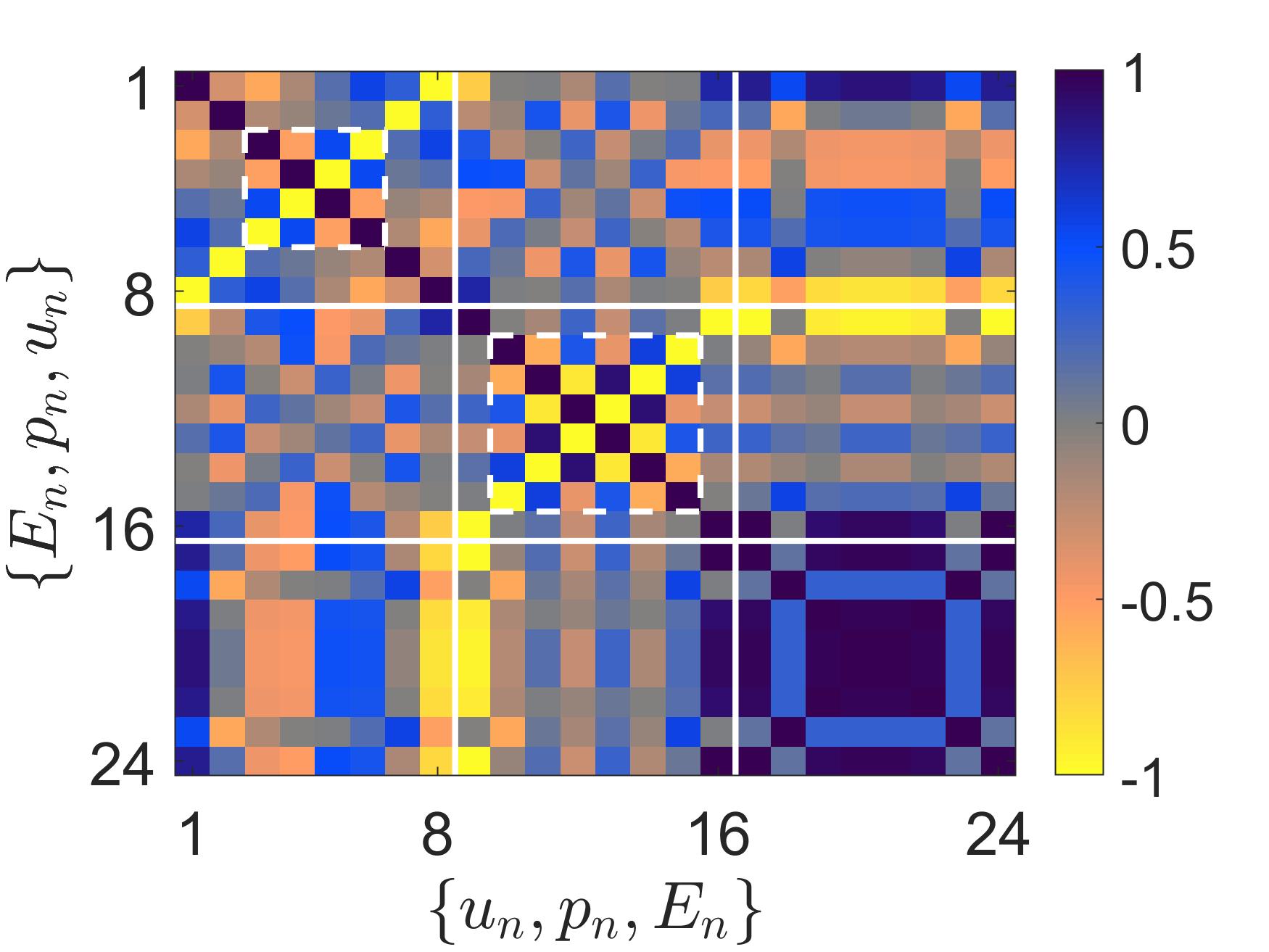}}
\subfigure[]{\label{fig:corrMp}
\includegraphics[trim=0cm 0cm 1.1cm 1cm,clip=true,width=0.48\textwidth]{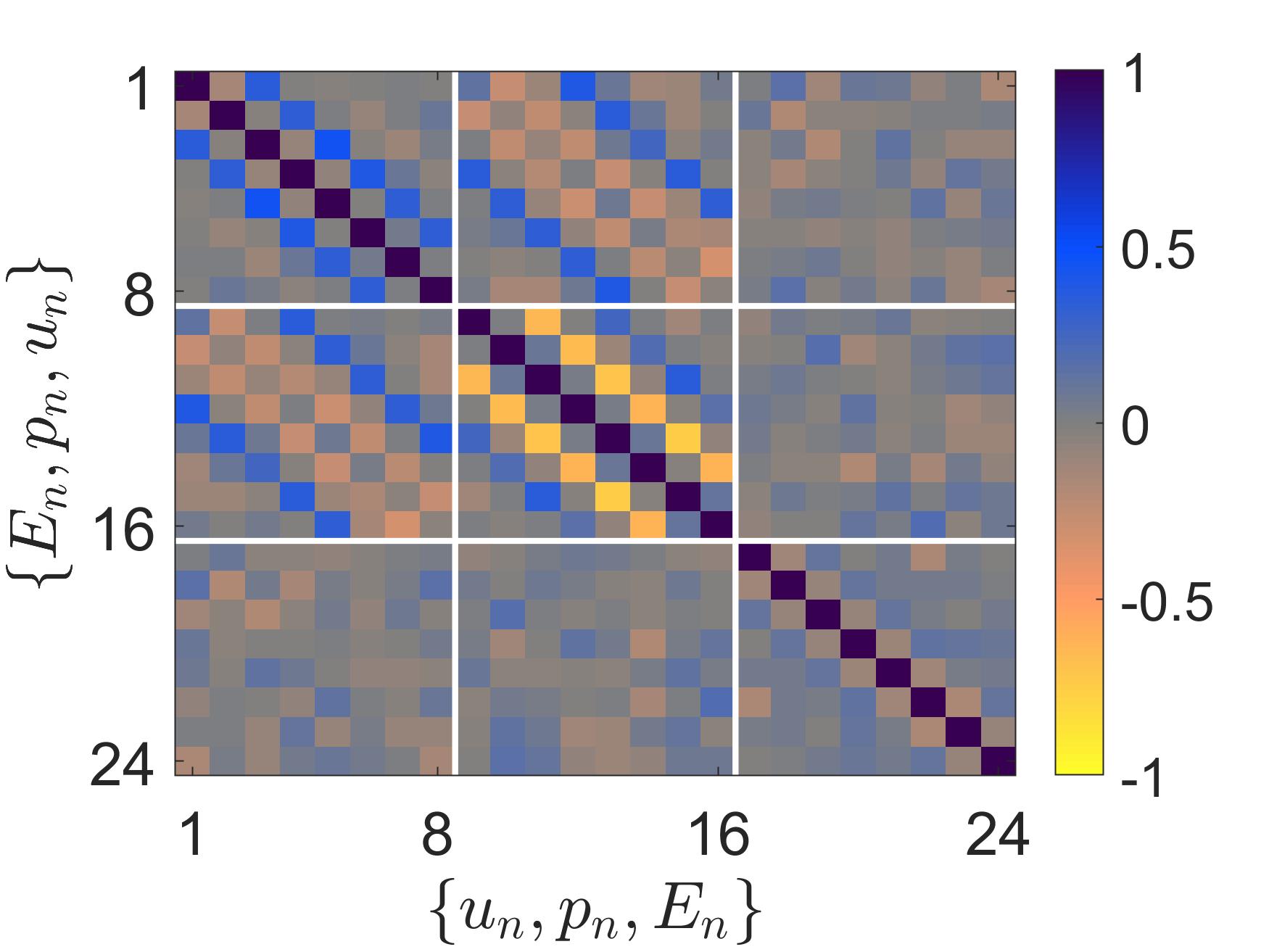}}
\caption{Pearson's correlation coefficients. (a) correlation matrix of discrete breather data of the dataset $X_{u,p,E}$. (b) correlation matrix of phonon data of the dataset $X_{u,p,E}$.}\label{fig:corrM}
\end{figure}

To increase the efficiency of the classification algorithms we consider two dimensionality reduction techniques, i.e., {\it Principal Component Analysis} (PCA) and {\it Locally Linear Embedding} (LLE) \cite{ML}. The objective is to find a low-dimensional representation of high-dimensional correlated datasets \eqref{eq:datasets} and identify clusters of localized and nonlocalized wave data. In Figure \ref{fig:corrM} we plot Pearson's correlation coefficients of the dataset $X_{u,p,E}$, i.e., the covariance matrix \eqref{eq:covM} of all pairs of particle sampled data variables, $u_n$, $p_n$ and $E_n$, divided by the product of their standard deviations. Pearson's correlation coefficient shows linear dependence (linear regression) between datasets, i.e., positive value indicates that positive increase in one variable yields also a positive increase in the other variable, while the negative value indicates a negative increase in the other variable. In addition, zero value indicates that the variables are  not correlated and the absolute value of coefficients indicates the correlation strength.

In Figure \ref{fig:corrMb} we consider only discrete breather data from $X_{u,p,E}$ and observe complex correlations between different particle dynamics data. Notice only positive correlations between particle energy values, see the right bottom square of Figure \ref{fig:corrMb}, and two squares indicated by dashed lines suggesting adjacent particle movement out of phase in Figure \ref{fig:Bsol}. On the other hand, in Figure \ref{fig:corrMp} we plot correlation matrix of phonon data of the dataset $X_{u,p,E}$. Notice very small correlations between sampled particle dynamics data of phonons. 

PCA is a data projection algorithm onto so called principal components that account for the largest amount of variance in sampled data. The algorithm is based on the eigen-decomposition of the covariance matrix
\begin{equation}\label{eq:covM}
C:= \frac{1}{N_{sim}-1} Y^T Y,
\end{equation}
where $Y:=X-\bar{X}$, $X\in \R^{N_{sim}\times K}$ is given dataset, such as $X_u$ or $X_{u,p}$, or $X_{u,p,E}$, and $\bar{X}$ is the matrix of the row-wise mean values $\bar{x}\in \R^{1\times K}$, i.e.,
\[
\bar{x}_j:=\frac{1}{N_{sim}}\sum_{i=1}^{N_{sim}}X_{ij}, \quad j=1,\dots,K, \quad 
\bar{X}:=\begin{pmatrix} 1 \\ \vdots \\ 1\end{pmatrix} \bar{x}.
\]
The covariance matrix \eqref{eq:covM} is symmetric and semi-positive definite. Thus, eigen-decomposition yields principal component orthogonal coordinate system with nonnegative eigenvalues which are squared singular values of the matrix $Y/\sqrt{N_{sim}-1}$. Lower dimensional representation $Z\in\R^{N_{sim} \times d}$, where $d\ll K$, is obtained by projecting data matrix $Y$ onto orthogonal coordinates with associated largest singular values. In our experiments we observe that with projection down to only two principal components, i.e., $d=2$, we are able to preserve more than $95\%$ of dataset's total variance what is found by computing variance of projected data onto each principal component.  

On the other hand, LLE is a dimensionality reduction technique that computes low dimensional, neighborhood preserving embeddings of high dimensional data. In contrast to PCA it does not rely on projections but evaluates linear relations between closest neighbors and constructs low-dimensional representation of data by attempting to preserve these relationships. Consider a dataset $X\in \R^{N_{sim}\times K}$. Mathematically \cite{ML} the algorithm is formulated as two optimization problems. In the first constrained optimization problem for each data instance $x^i:= X_i^T\in\R^{K}$, $i=1,\dots,N_{sim}$, i.e., $x^i$ is the transpose of the $i^{th}$ row of the matrix $X$, we search for its $k$ closest neighbors $x^j:=X_j^T$ and weight values $w_{i,j}$ which give the best linear representation for the data instance $x^i$ of $k$ closest neighbors $x^j$ measured by the square distance error, i.e.,
\[
\min_{w} \sum_{i=1}^{N_{sim}}\bigg| x^i - \sum_{j=1}^{N_{sim}}w_{i,j}x^{j}\bigg|^2,
\]
subject to the following constraints:
\[
\begin{cases}
w_{i,j}=0, & \mbox{if $x^j$ is not close neighbor of $x^i$ or $j=i$},\\
\displaystyle \sum_{j=1}^{N_{sim}}w_{i,j}=1, & \mbox{for all} \, i=1,\dots , N_{sim},
\end{cases}
\]
where the second constraint imposes invariance under translation.

Obtained matrix $W\in\R^{N_{sim}\times N_{sim}}$ of all weight values $w_{i,j}$ contains the local linear relationships between the data instances. We assume that localized wave data in high-dimensional space will be closer and further away from the linear nonlocalized data points. In our experiments performing LLE we considered twenty neighbors $(k=20)$ for each data point. In the second unconstrained optimization problem we minimize the square distance error to find low-dimensional representation $z^i\in \R^d$ of $x^i\in\R^{K}$, where $d=2 \ll K$ in our results, while preserving local linear relationships, i.e.,
\[
\min_{z} \sum_{i=1}^{N_{sim}}\bigg| z^i - \sum_{j=1}^{N_{sim}}w_{i,j}z^{j}\bigg|^2.
\]
Once the low-dimensional dataset $Z\in\R^{N_{sim} \times d}$, i.e., $Z_i:={z^{i}}^T$, is obtained, we subtract row-wise mean values and scale columns to unit variance before building classification algorithms. 

\begin{figure}[p]
\centering 
\subfigure[]{\label{fig:PCAW}
\includegraphics[trim=0cm 0cm 0cm 0cm,clip=true,width=0.475\textwidth]{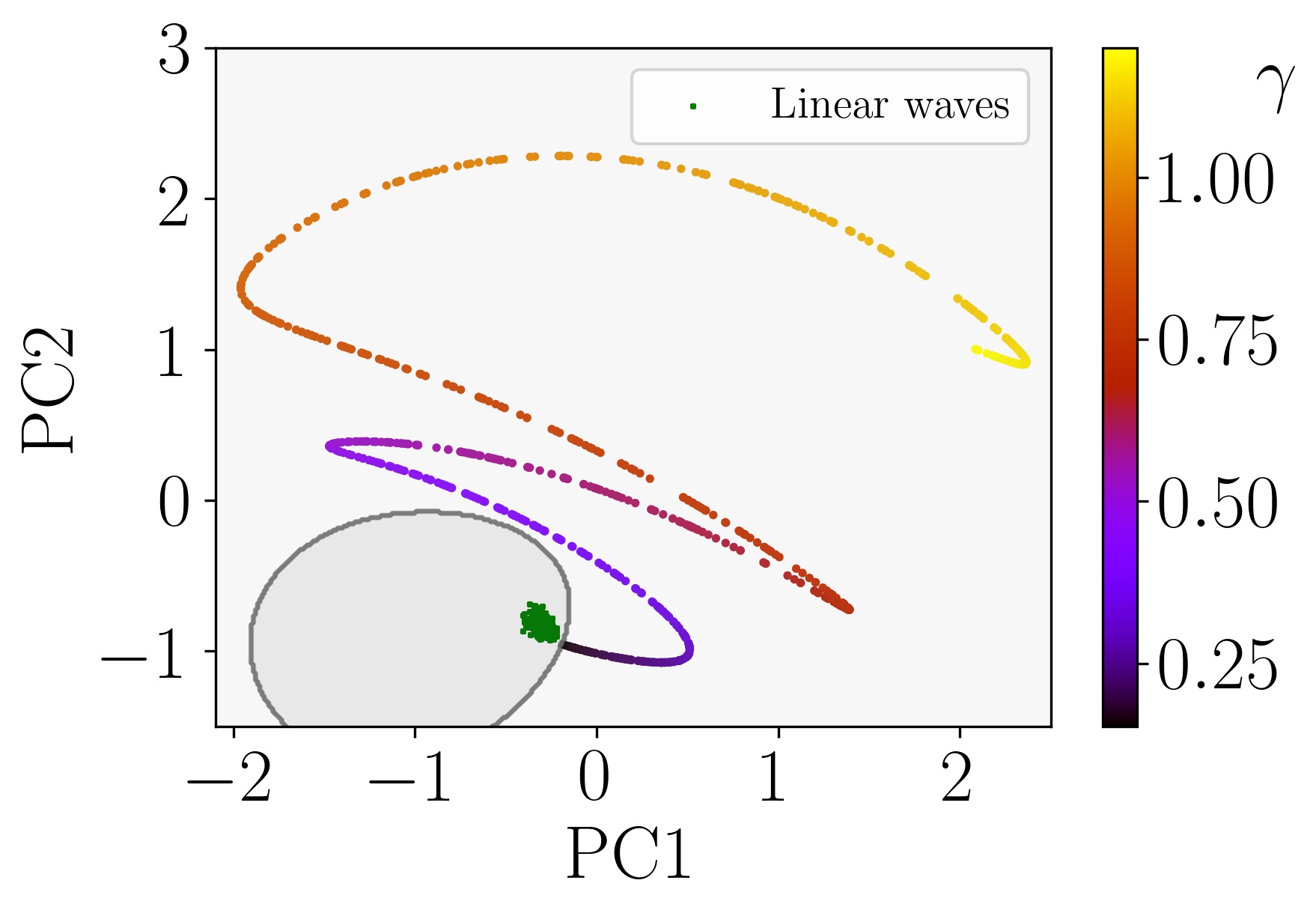}}
\subfigure[]{\label{fig:LLEW}
\includegraphics[trim=0cm 0cm 0cm 0cm,clip=true,width=0.475\textwidth]{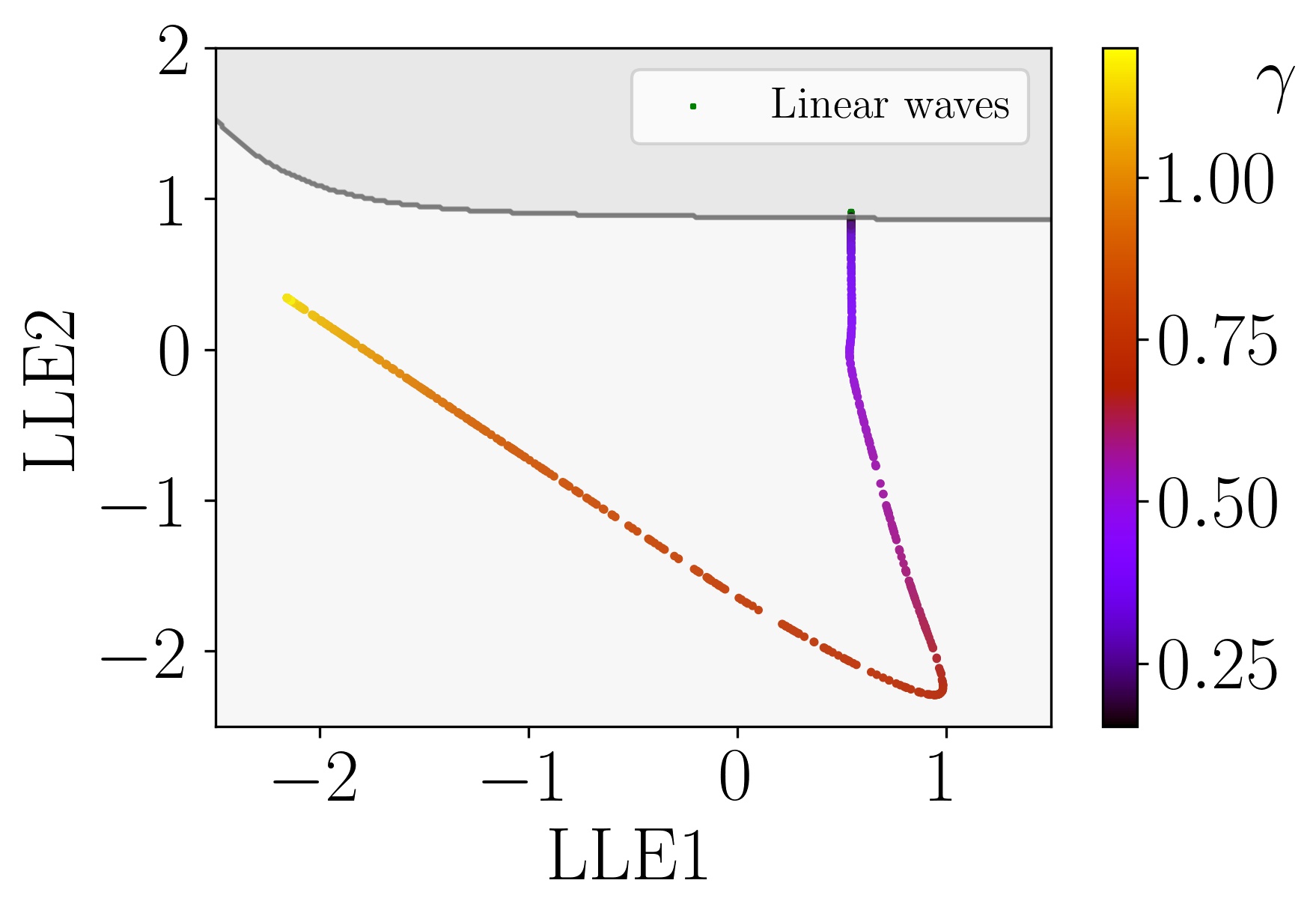}}
\subfigure[]{\label{fig:PCAWP}
\includegraphics[trim=0cm 0cm 0cm 0cm,clip=true,width=0.475\textwidth]{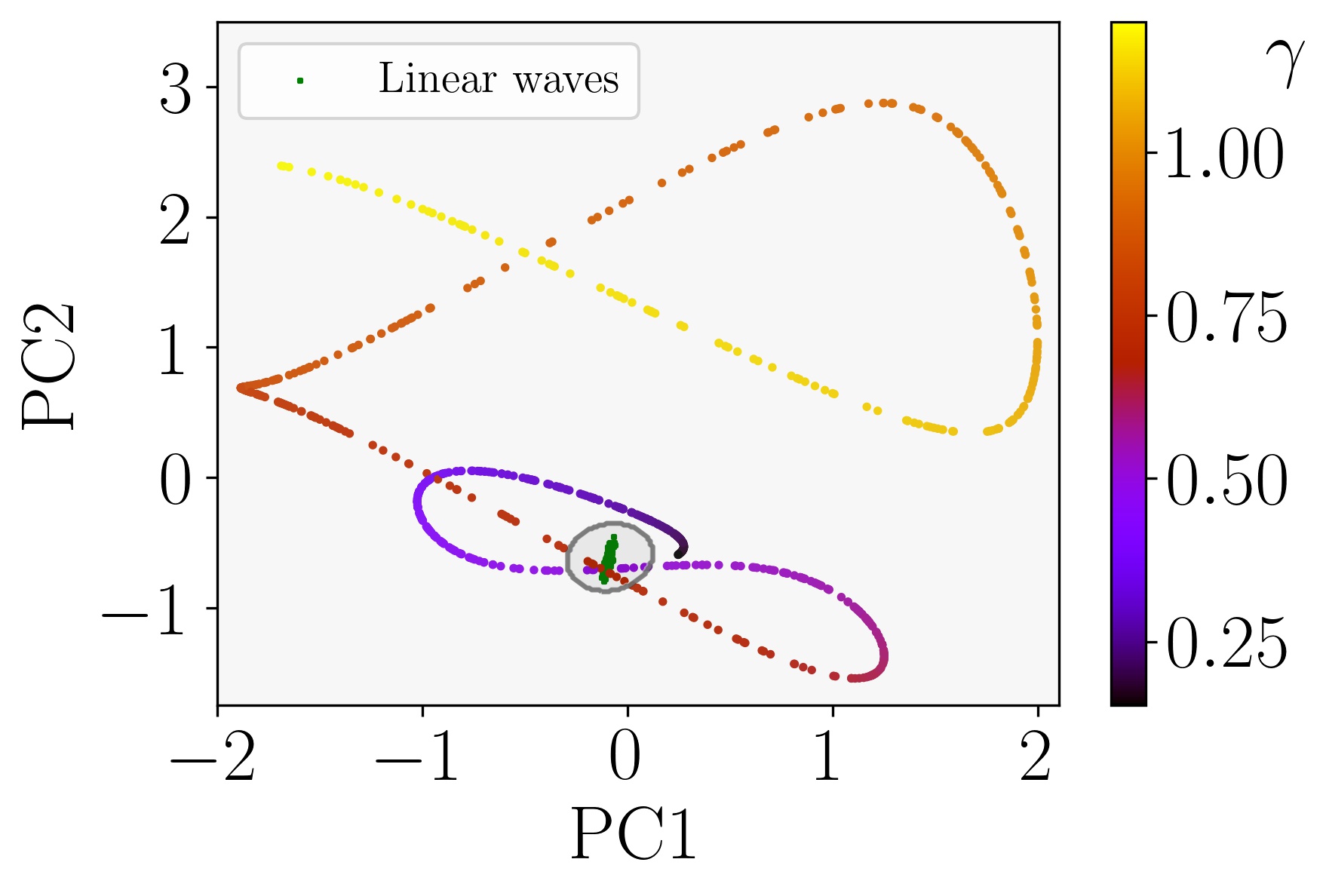}}
\subfigure[]{\label{fig:LLEWP}
\includegraphics[trim=0cm 0cm 0cm 0cm,clip=true,width=0.475\textwidth]{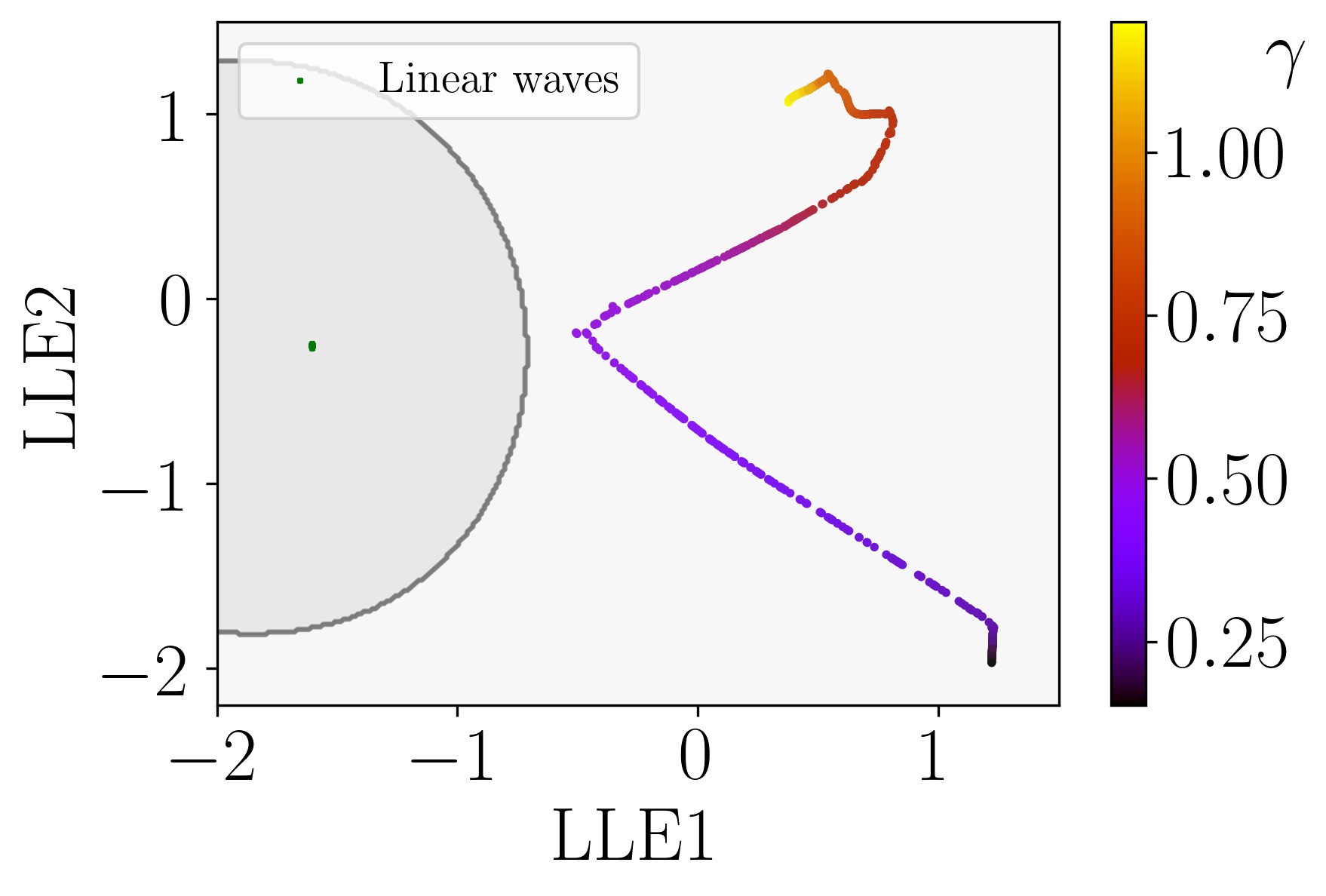}}
\subfigure[]{\label{fig:PCAX}
\includegraphics[trim=0cm 0cm 0cm 0cm,clip=true,width=0.475\textwidth]{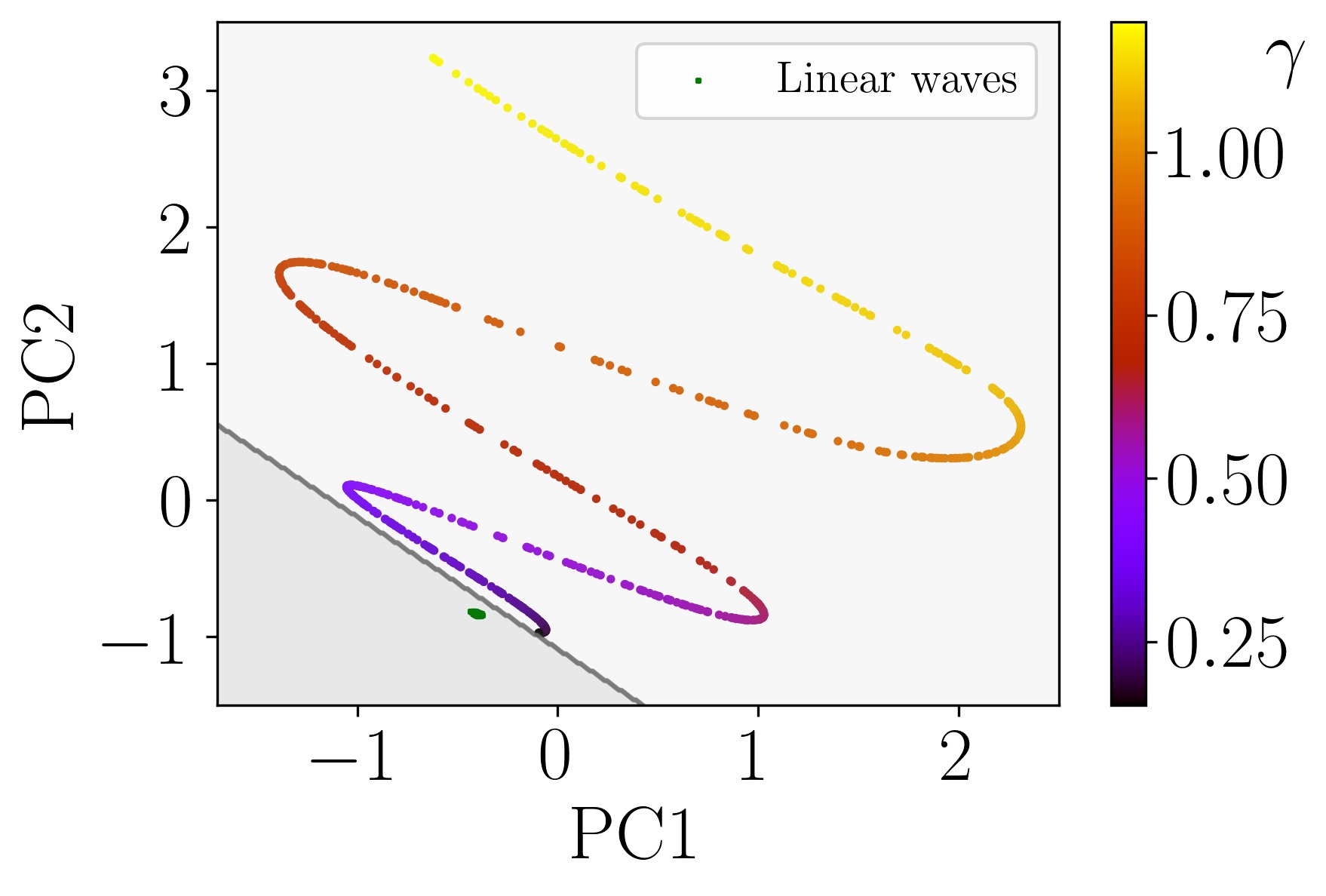}}
\subfigure[]{\label{fig:LLEX}
\includegraphics[trim=0cm 0cm 0cm 0cm,clip=true,width=0.475\textwidth]{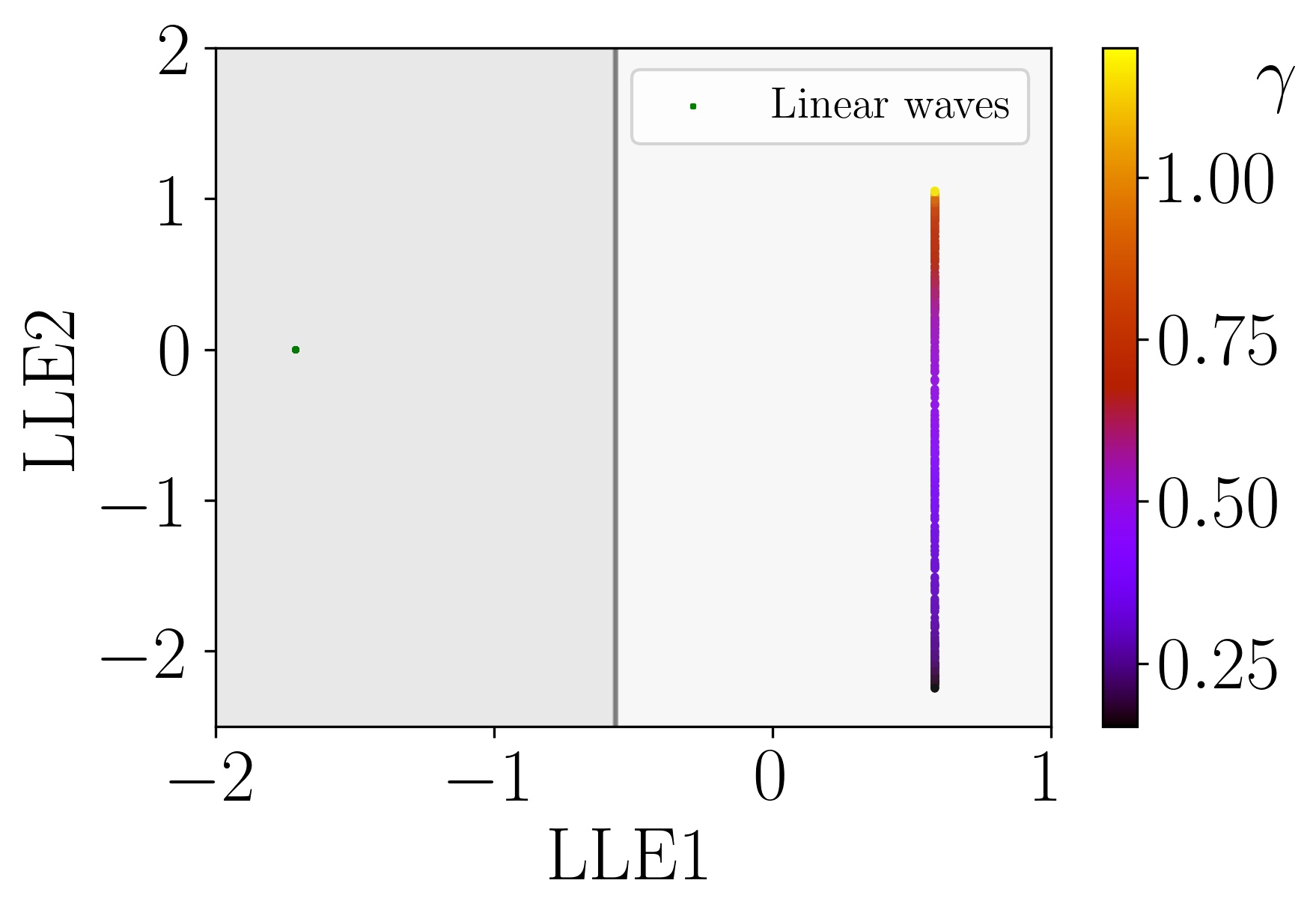}}
\caption{Dimensionality reduction with PCA and LLE of three datasets $X_{u}$, $X_{u,p}$ and $X_{u,p,E}$ and decision boundaries \eqref{eq:db} of SVC indicated by solid gray curve which separates two prediction regions. Colorbars indicate $\gamma$ values of the pattern \eqref{eq:patern} used to obtain DB solution data. (a) PCA of $X_{u}$ with nonlinear SVC, $\lambda=20$. (b) LLE of $X_{u}$ with nonlinear SVC, $\lambda=80$. (c) PCA of $X_{u,p}$ with nonlinear SVC, $\lambda=80$. (d) LLE of $X_{u,p}$ with nonlinear SVC, $\lambda=0.01$. (e) PCA of $X_{u,p,E}$ with linear SVC, $\lambda=1$. (f) LLE of $X_{u,p,E}$ with linear SVC, $\lambda=0.01$.}\label{fig:dimred}
\end{figure}

In Figure \ref{fig:dimred} we illustrate dimension reduction of all three datasets $X_{u}$, $X_{u,p}$ and $X_{u,p,E}$ applying both dimensionality reduction techniques PCA and LLE, respectively. Colorbars in Figure \ref{fig:dimred} indicate $\gamma$ values of the pattern \eqref{eq:patern} used to obtain particular DB solution data. Apart from the PCA projection of the datasets $X_{u,p}$, see Figure \ref{fig:PCAWP}, and $X_{p}$, not shown, the $\gamma$ values can be traced along the continuous curve in two dimensions. Interestingly, in Figure \ref{fig:LLEX} the curve is a straight line parallel to $y$-axis, which was observed for all datsets containing energy values, i.e., $X_E$, $X_{u,E}$, $X_{p,E}$ and $X_{u,p,E}$. 

In Figures \ref{fig:PCAW}, \ref{fig:PCAWP} and \ref{fig:PCAX} we projected datasets down onto two principal components $(PC1,PC2)$, while in Figures \ref{fig:LLEW}, \ref{fig:LLEWP} and \ref{fig:LLEX} we show local data linear embedding in two-dimensional $(LLE1,LLE2)$ coordinates. Notice that linear wave data is hardly separated from nonlinear localized wave data in Figures \ref{fig:PCAW}, \ref{fig:LLEW} and \ref{fig:PCAWP}. For the dataset $X_{u,p}$ Figure \ref{fig:LLEWP} suggests that LLE is superior over PCA for the classification problem. Locally linear embeddings of the dataset $X_{u,p,E}$ (shown) and datasets $X_E$, $X_{u,E}$ and $X_{p,E}$ (not shown) are able to produce the largest separation between two types of wave data. Thus, information of particle displacements and momenta alone may not be sufficient to build an optimal classifier for detecting ILMs, e.g., in numerical simulations, see Section \ref{sec:Applications}.     

\subsection{Classification algorithms}

Linear and nonlinear kernel {\it Support Vector Machine} \cite{ML,Bishop} classifiers (SVCs) are trained for lattice wave data classification, see Figure \ref{fig:dimred}. SVCs are supervised learning algorithms that construct a decision boundary given by the equation:
\begin{equation}\label{eq:db}
w^T \phi(z) + b  = 0, \quad z\in\R^d,
\end{equation}
with the largest distance, also called a margin, to the nearest training-data point of each data class. The function $\phi: \R^d \to \R^D$ maps $z$ into higher $D$-dimensional feature space $(D \geq d)$. $w\in\R^{D}$ and $b\in\R$ are unknown parameters to be determined such that the prediction function $\sgn\left( w^T \phi(z) + b \right)$ for two class classification problem is correct for most of the training-data instances $z=z^i:=Z_i^T$, where $Z\in\R^{N_{sim}\times d}$. Special case includes linear classifier with $\phi(z)=z$. Then, e.g., in two dimensions equation \eqref{eq:db} defines a line as decision boundary, see Figures \ref{fig:PCAX} and \ref{fig:LLEX}. 

SVC algorithm is formulated as a constrained optimization problem \cite{Bishop}:
\begin{equation}\label{eq:SVCa}
\min_{w,b,\zeta} \left(
\frac{1}{2}w^Tw + \lambda \sum_{i=1}^{N_{sim}}\zeta_i 
\right)
\end{equation}
with respect to a subject to the constraints
\begin{equation}\label{eq:SVCb}
\begin{cases}
y_i \left( w^T \phi(z^i) + b \right) \geq 1 - \zeta_i,\\
\zeta_i \geq 0,
\end{cases}
\end{equation}
for all $i=1,\dots,N_{sim}$, where $y_i\in\{-1,1\}$ is two class, i.e., localized and nonlocalized wave, label value for the data instance $z^{i}$. Constant $\lambda \geq 0$ is the regularization hyperparameter which makes possible more flexible soft margine classification through the introduction of nonnegative variables
\[
\zeta_i := \max(0, 1 - y_i \left( w^T \phi(z^i) + b \right))
\]
which allows for some data instances to be misclassified, i.e.,
\begin{itemize}
\item if $\zeta_i=0$, then the data point is correctly classified;
\item if $0 < \zeta_i < 1$, then the data point lies inside the margin on the correct side of the decision boundary \eqref{eq:db};
\item if $\zeta_i = 1$, then the data point lies on the decision boundary \eqref{eq:db};
\item if $\zeta_i > 1$, then the data point lies on the incorrect side of the decision boundary \eqref{eq:db} and is misclassified.
\end{itemize}
 
In order to solve the optimization problem \eqref{eq:SVCa}--\eqref{eq:SVCb}, especially when $D \gg 1$ or even infinite, we define Lagrange multipliers $a_i \geq 0$ and $\mu_i\geq 0$ of the constraints \eqref{eq:SVCb}, respectively, and derive a dual quadratic programming problem \cite{Bishop,CC11}
\begin{equation}\label{eq:SVCa_dual}
\min_{a} \left(
\frac{1}{2} 
\sum_{i=1}^{N_{sim}} \sum_{j=1}^{N_{sim}} a_i a_j y_i y_j \mathcal{K}(z^i,z^j) - \sum_{i=1}^{N_{sim}} a_i 
\right),
\end{equation}
subject to the following constraints:
\begin{equation}\label{eq:SVCb_dual}
\begin{cases}
\displaystyle \sum_{i=1}^{N_{sim}} y_i a_i = 0,\\
0 \leq a_i \leq \lambda, & \mbox{for all } i=1,\dots,N_{sim},
\end{cases}
\end{equation}
where $\mathcal{K}(z^i,z^j):=\phi(z^i)^T\phi(z^j)$ is symmetric Kernel function which can be defined without explicit knowledge of the transformation $\phi$. Note that the Lagrange multipliers $\mu_i\geq 0$ do not appear in the optimization problem \eqref{eq:SVCa_dual}--\eqref{eq:SVCb_dual} but imply that $a_i \leq \lambda$. In addition, if $0< a_i < \lambda$, then $\zeta_i=0$, and if $a_i = \lambda$, then $\zeta_i > 0$, while the zero values $a_i=0$ do not contribute to the prediction, since the optimal $w$ of the problem \eqref{eq:SVCa}--\eqref{eq:SVCb} is given by
\[
w = \sum_{i=1}^{N_{sim}} y_i a_i \phi(z^i)
\]
and the prediction function is
\[
\sgn\left( w^T \phi(z) + b \right) = 
\sgn\left( \sum_{i=1}^{N_{sim}} y_i a_i \mathcal{K}(z^i,z) + b \right) 
\]
where $b$ can be determined noting that for all $0< a_i < \lambda$ we have that $\zeta_i=0$ which implies that $y_i \left( w^T \phi(z^i) + b \right)=1$. Thus,
\[
y_i \left(  \sum_{\substack{j=1 \\ 0< a_j < \lambda }}^{N_{sim}} y_j a_j \mathcal{K}(z^j,z^i) + b \right)=1 
\quad \implies \quad
b = y_i  -  \sum_{\substack{j=1 \\ 0< a_j < \lambda }}^{N_{sim}} y_j a_j \mathcal{K}(z^j,z^i)  
\]
where numerically stable solution is obtained by averaging over all indexes $i$ for which $0< a_i < \lambda$.

In addition to linear kernel SVCs, $\mathcal{K}(z^i,z^j):={z^i}^T z^j$, the Gaussian {\it Radial Basis Function} (RBF) kernel, $\mathcal{K}(z^i,z^j):=\exp(-\alpha |z^i-z^j|^2)$ where $\alpha>0$, is used in nonlinear SVCs, see Figures \ref{fig:PCAW}--\ref{fig:LLEWP}. It showed improved performance over other nonlinear kernels, such as polynomial and hyperbolic tangent \cite{ML}. Soft margin SVCs \eqref{eq:SVCa_dual}--\eqref{eq:SVCb_dual} are considered where optimal hyperparameter values, such as $\lambda$ and $\alpha$, can be found with {\it grid search} technique \cite{ML}. We set problem scaled $\alpha$ value, i.e., $\alpha=1/d=0.5$, while all $\lambda$ values are found and indicated in Figure's \ref{fig:dimred} caption.

To evaluate the performance of linear and nonlinear classifiers for each dataset $\mathrm{K}$-fold cross-validation \cite{ML} was considered with $0.7$ training and testing data proportion, i.e., $70\%$ of random data samples of both wave types from $N_{sim}$ simulations were considered for training and cross-validation, while remaining $30\%$ of the data was used for testing. In our experiments we considered $\mathrm{K}=5$. The {\it precision} and {\it recall} scores \cite{ML} were recorded on cross-validation and testing datasets. The precision is a metric of accuracy of the positive (localized wave) predictions while the recall is a metric of positive instances that are correctly detected by the classifier.

\begin{table}[t]
	\begin{center}
		\begin{tabular}{|c||c|c||c|c||c|}
			\hline
			\multirow{2}{*}{} & \multicolumn{2}{c||}{{Linear kernel}} &
			\multicolumn{2}{c||}{{RBF kernel}} & \\
			\cline{2-5}
			& {PCA} & {LLE} & {PCA} & {LLE} & \\			
			\specialrule{.1em}{.05em}{.05em} 
			{Precision} (validation set) & 0.9985 & 0.9623 & 1 & 0.9953
			& \multirow{4}{*}{$X_u$} \\
			{Recall} (validation set) & 0.9533 & 0.9478 & 0.9998 & 0.9354 & \\
			\cline{1-5}
			{Precision} (testing set) & 0.9984 & 0.9658 & 1 & 0.9965 & \\
			{Recall} (testing set) & 0.9516 & 0.9487 & 0.9992 & 0.9370 & \\
			\specialrule{.1em}{.05em}{.05em} 
			{Precision} (validation set) & 0.7947 & 0.9927 & 1 & 1
			& \multirow{4}{*}{$X_{u,p}$} \\
			{Recall} (validation set) & 
0.9544 & 0.9993 & 0.9754 & 0.9990 & \\
			\cline{1-5}
			{Precision} (testing set) & 0.7978 & 0.9928 & 1 & 0.9999 & \\
			{Recall} (testing set) & 
0.9519 & 0.9999 & 0.9758 & 0.9991 & \\
						
			\specialrule{.1em}{.05em}{.05em} 
			{Precision} (validation set) & 1 & 1 & 1 & 1 & \multirow{4}{*}{$X_{u,p,E}$} \\
			{Recall} (validation set) & 1 & 1 & 1 & 1 & \\
			\cline{1-5}
			{Precision} (testing set) & 1 & 1 & 1 & 1 & \\
			{Recall} (testing set) & 1 & 1 & 1 & 1 & \\
			\hline
		\end{tabular}
	\end{center}
	\caption{Precision and recall metrics averaged over $100$ random train-test-splits of different classification algorithms, dimensionality reduction techniques and datasets.}
	\label{tab:Precision_recall}
\end{table}

Averaged precision and recall values over $100$ random  {\it train-test-splits} of datasets $X_{u}$, $X_{u,p}$ and $X_{u,p,E}$ with $0.7$ data proportion are listed in Table \ref{tab:Precision_recall}. For each train-test-split we compute precision and recall scores for the testing data and validation dataset obtained from the $\mathrm{K}$-fold of training data. Notice that in most cases recall values are smaller than precision values which is attributed to the fact that more localized wave data is classified as linear opposed to linear waves being classified as localized. In addition to mean values we also computed (not shown) variance values which in all cases, if not exactly zero, did not exceed $10^{-2}$ value.

Examining Table \ref{tab:Precision_recall} we observe exceptionally good results except for the PCA of the dataset $X_{u,p}$ when linear SVC is used, what is also true for the dataset $X_p$. That is also evident in Figure \ref{fig:PCAWP}, where it is easy to see that nonlinear waves can not be separated from linear waves by a straight line. In that case the use of nonlinear SVC is required. Very good results are demonstrated by PCA and LLE dimensionality reduction of the dataset $X_{u,p,E}$, including all other datasets containing energy values, either with linear or nonlinear SVC. In this case for simplicity we advocate the use of linear SVC. Precision and recall values equal to one state that none of the data samples were incorrectly classified by SVC. As follows we consider linear SVC for the dataset $X_{u,p,E}$ and nonlinear classifier for the datasets $X_{u}$ and $X_{u,p}$, which is illustrated in Figure \ref{fig:dimred} where the decision boundaries \eqref{eq:db} are indicated by solid gray curves clearly separating two prediction regions where darker region indicates prediction region of nonlocalized waves, while the lighter region indicates prediction region of localized waves. Arguments for using the dataset $X_{u,p,E}$ instead of just $X_{E}$ will be stated in the following section.

In summary, Table \ref{tab:Precision_recall} shows that the classifier is not sensitive to random train-test-split of the datasets and that for the dataset $X_{u,p,E}$ SVC is essentially a perfect classifier for both dimensionality reduction techniques. To further differentiate the classifiers we consider their classification performance on imperfect data, see the following section.           

\subsection{Classification of imperfect data}

So far we have considered datasets of perfectly sampled data of linear and localized nonlinear waves. With application in mind, i.e., detecting ILMs in numerical simulations, we consider additional datasets $X_{u,p,E}^{n_{off}}$ of localized discrete breather data from $N_{sim}=1000$ simulations with randomly chosen $\gamma$ values from the uniform distribution $\mathcal{U}(0.15,1.2)$. We define index $n_{off}$, where $n_{off}=0,1,\dots,8$, which indicates the shift in particles from which the data is sampled, i.e., the data is sampled from particles $q_{m+n_{off}}$, where particle $q_m$ is the first particle in Figure \ref{fig:Bsol} identified with red dot. Thus, the sampled datasets $X_{u,p,E}^{n_{off}}$ consist of partially linear and nonlinear wave data. 

\begin{figure}[t]
\centering 
\subfigure[]{\label{fig:dataoffA}
\includegraphics[trim=0cm 0cm 0cm 0cm,clip=true,width=0.48\textwidth]{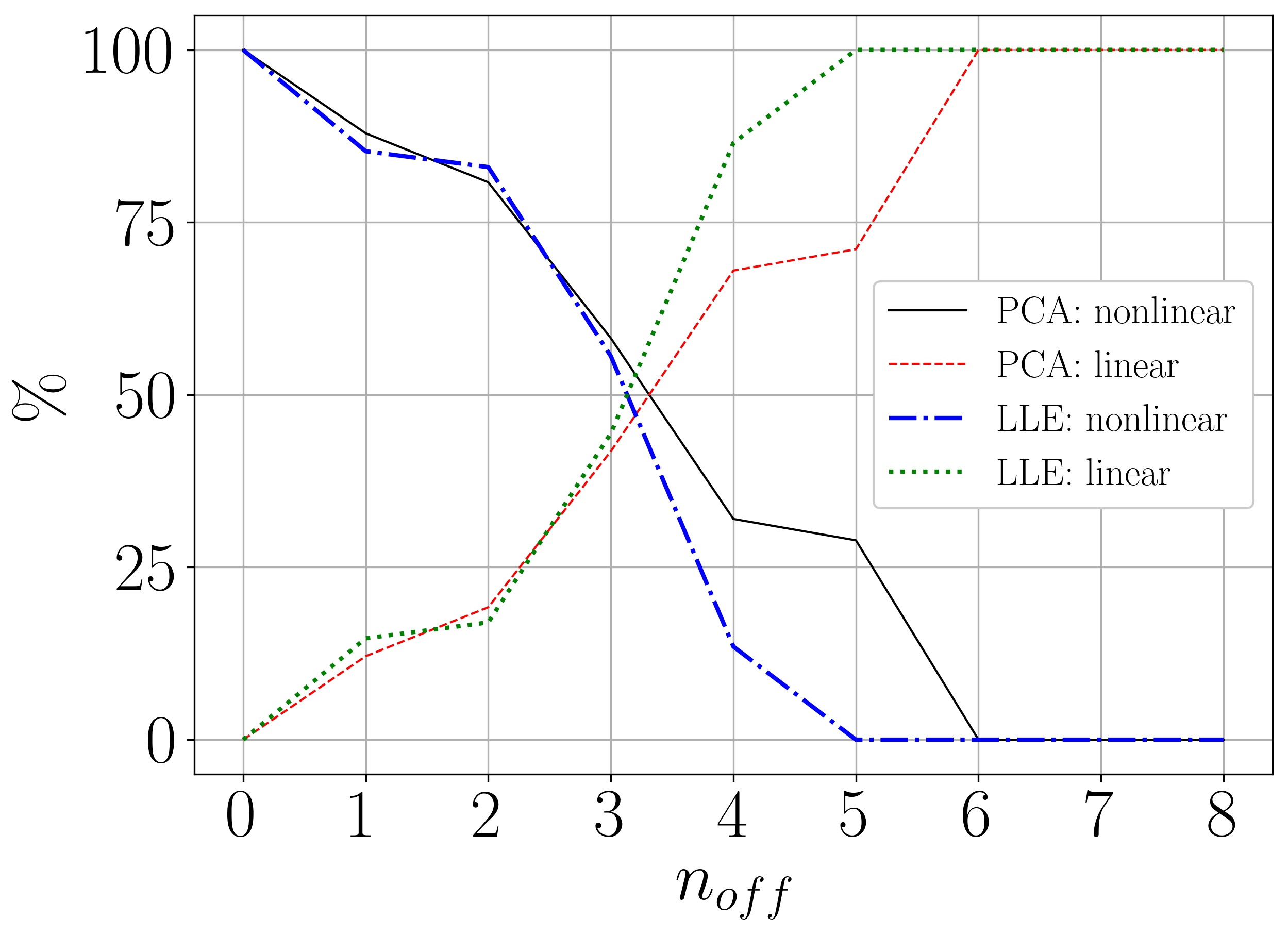}}
\subfigure[]{\label{fig:dataoffB}
\includegraphics[trim=0cm 0cm 0cm 0cm,clip=true,width=0.48\textwidth]{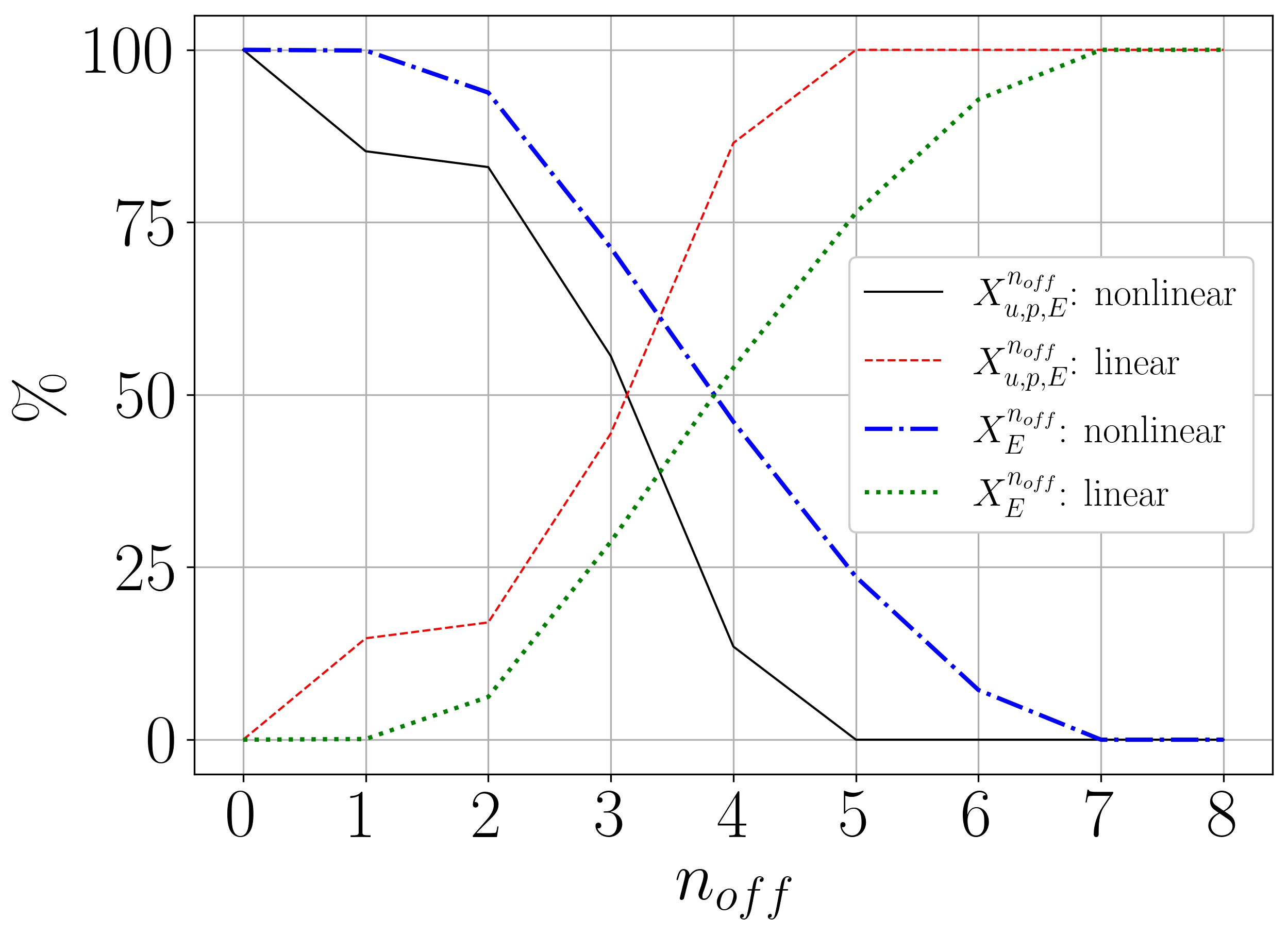}}
\caption{Percentages of classified imperfect localized wave data. (a) linear SVC with dimensionality reduction methods PCA and LLE of the datasets $X_{u,p,E}^{n_{off}}$. (b) linear SVC with dimensionality reduction method LLE of the datasets $X_{u,p,E}^{n_{off}}$ and $X_{E}^{n_{off}}$.}\label{fig:dataoff}
\end{figure}

We classify samples from the datasets $X_{u,p,E}^{n_{off}}$ using linear SVC and both dimensionality reduction methods PCA and LLE. In Figure \ref{fig:dataoffA} we plot the percentage of wave data being classified as localized wave or linear wave depending on the shift index $n_{off}$. When $n_{off}=0$, the data consists of purely nonlinear wave data and we can see $100\%$ identification of it. On the other hand, when $n_{off}=8$ the dataset does not contain information of nonlinear waves and we see $100\%$ identification of linear wave data. Both classifiers demonstrate gradual decrease in localized wave identification as $n_{off}$ value increases. Notice that the curve of LLE drops faster to zero compared to PCA. This has an impact on detection region size, i.e., SVC with PCA produces wider detection regions of localization as indicated by numerical simulations of Section \ref{sec:Applications}.

In addition, in Figure \ref{fig:dataoffB} we compare two essentially perfect linear SVCs trained on the datasets $X_{u,p,E}$ and $X_{E}$ with LLE dimensionality reduction applied to the datasets $X_{u,p,E}^{n_{off}}$ and $X_{E}^{n_{off}}$. For both methods we observe desirable trend but the classifier based on the data consisting only of the energy data will produce much wider detection regions of the localization. Thus, we postulate that energy data \eqref{eq:En} is important to be included in datasets for training ILM classifiers but may be not sufficient.

\section{Applications}\label{sec:Applications}

In this section we apply classification algorithms developed above to detect localized DBs of the one-dimensional crystal lattice model \eqref{eq:Hamilt}. We consider sliding window object detection method by sliding a window of eight particle size at each time step over the computational data. To detect region of a localized wave we define a discrete density function $\rho_n$, $n=1,\dots,N$, which takes a positive value if data at particular eight particles is classified as being of localized wave, otherwise the density value is equal to zero. In the following Figures \ref{fig:ApplA} and \ref{fig:ApplBC} we identify all particles by red dots where the localization is detected.

For the first example we consider a lattice of $N=64$ particles and excite two stationary breathers using momenta pattern \eqref{eq:patern} with $\gamma=0.55$ and $\gamma=0.35$ at particles with indices $(16,17,18,19)$ and $(44,45,46,47)$, respectively. We solve the Hamiltonian system \eqref{eq:q}--\eqref{eq:p} until time $T_{end}=1$ and detect the regions of localization, see Figure \ref{fig:ApplA}.

\begin{figure}[t]
\centering 
\subfigure[]{\label{fig:ApplA_1}
\includegraphics[trim=1.4cm 0.5cm 2cm 3.5cm,clip=true,width=0.48\textwidth]{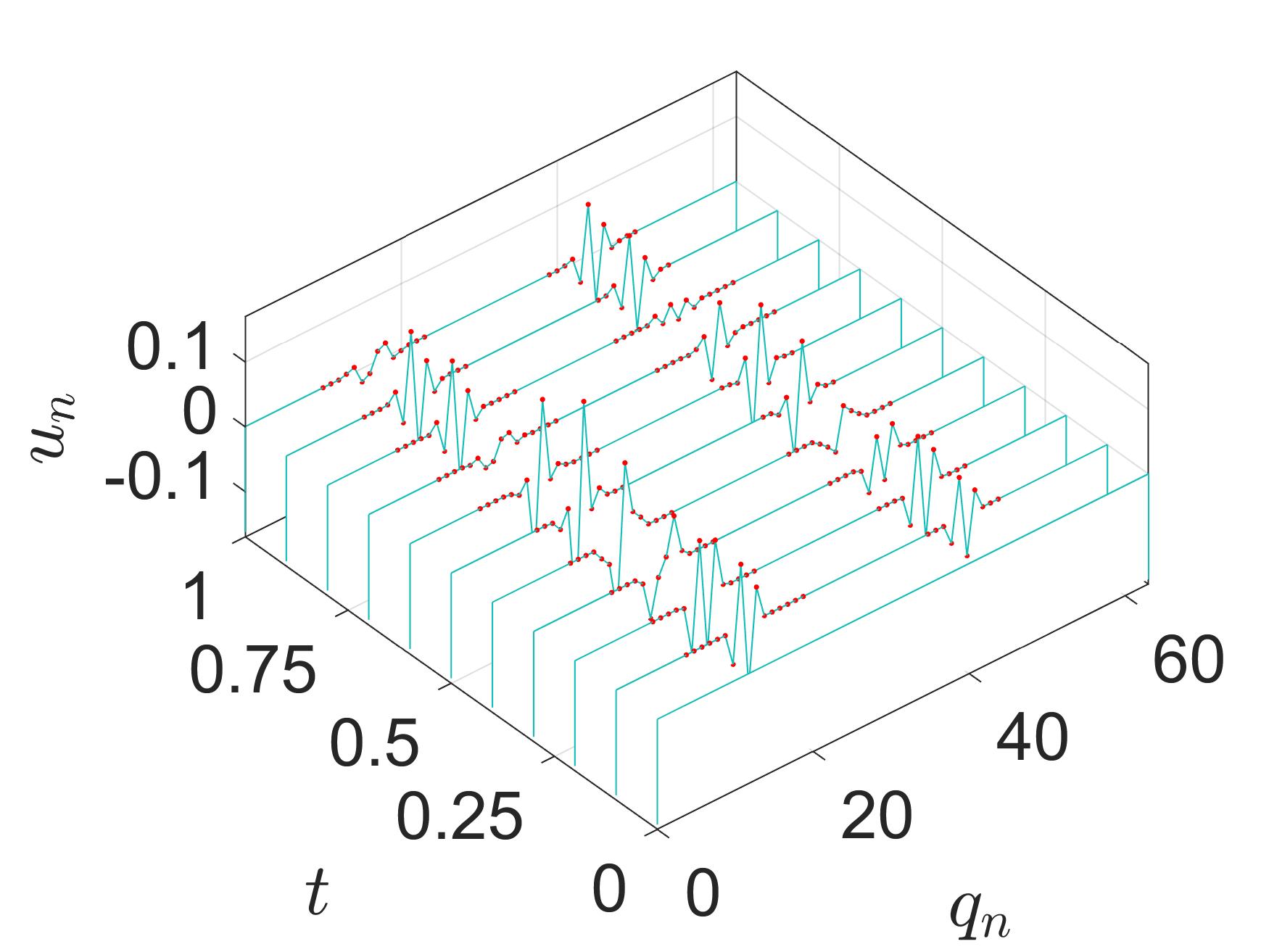}}
\subfigure[]{\label{fig:ApplA_2}
\includegraphics[trim=1.4cm 0.5cm 2cm 3.5cm,clip=true,width=0.48\textwidth]{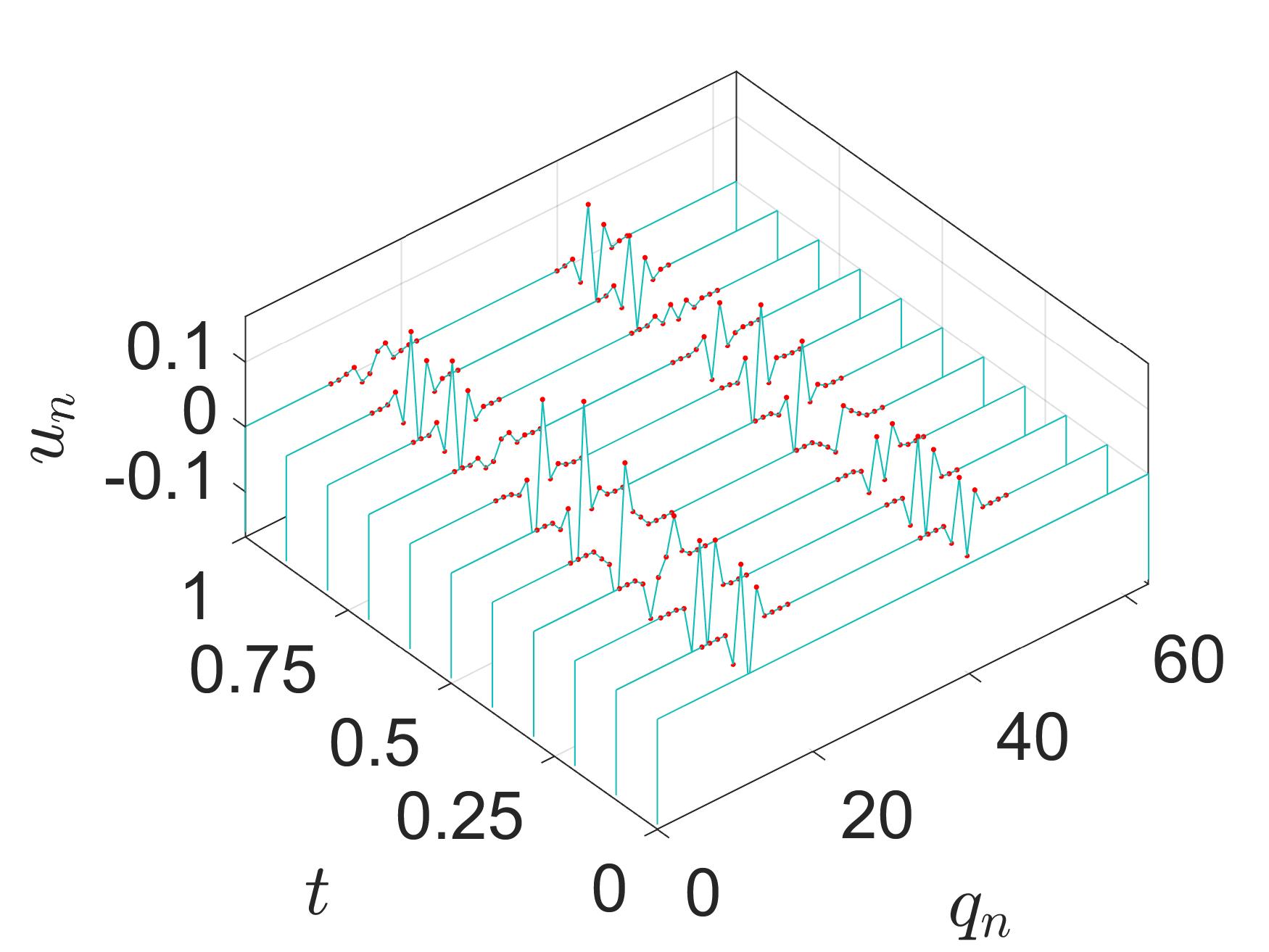}}
\subfigure[]{\label{fig:ApplA_3}
\includegraphics[trim=1.4cm 0.5cm 2cm 3.5cm,clip=true,width=0.48\textwidth]{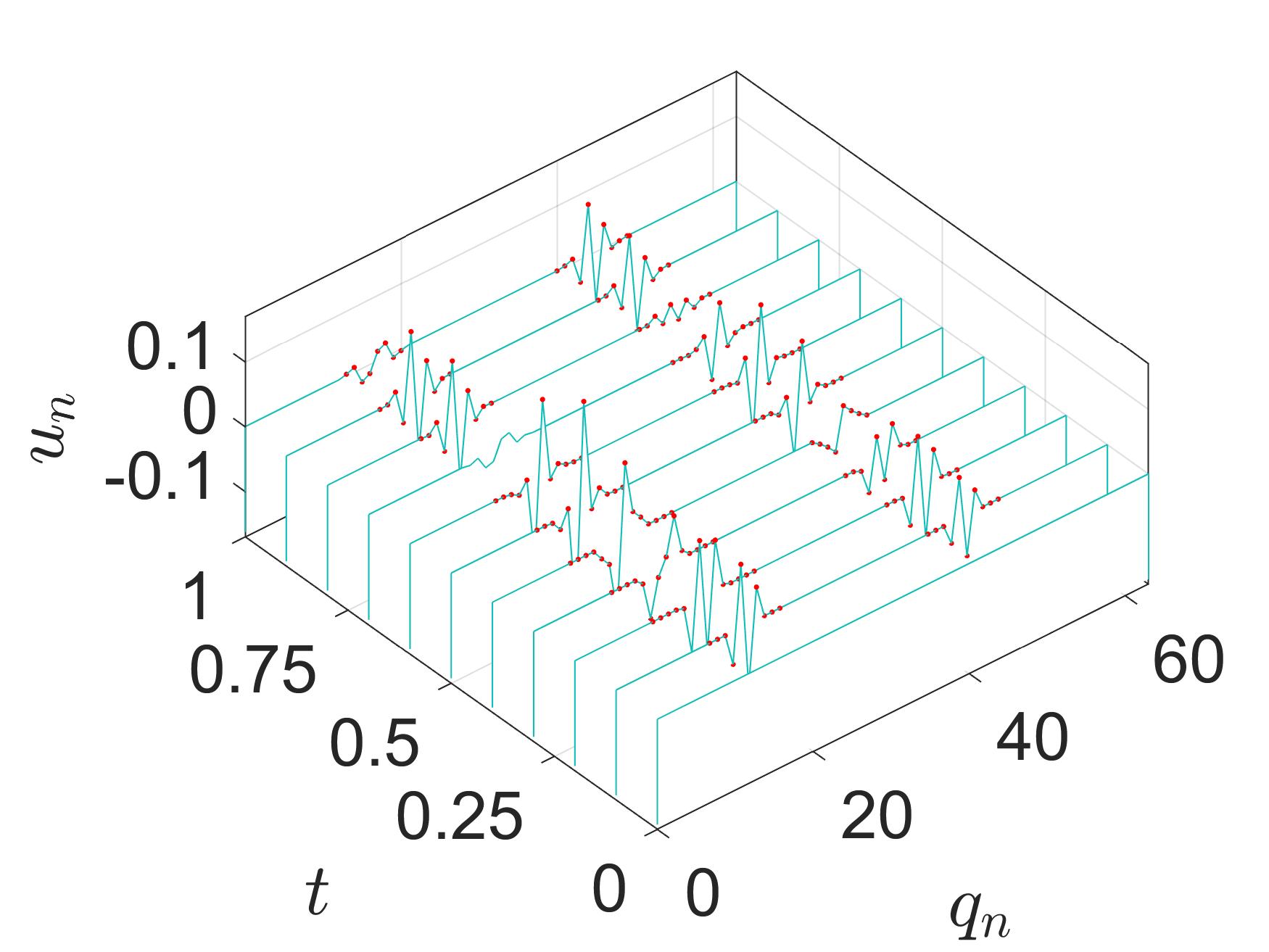}}
\subfigure[]{\label{fig:ApplA_4}
\includegraphics[trim=1.4cm 0.5cm 2cm 3.5cm,clip=true,width=0.48\textwidth]{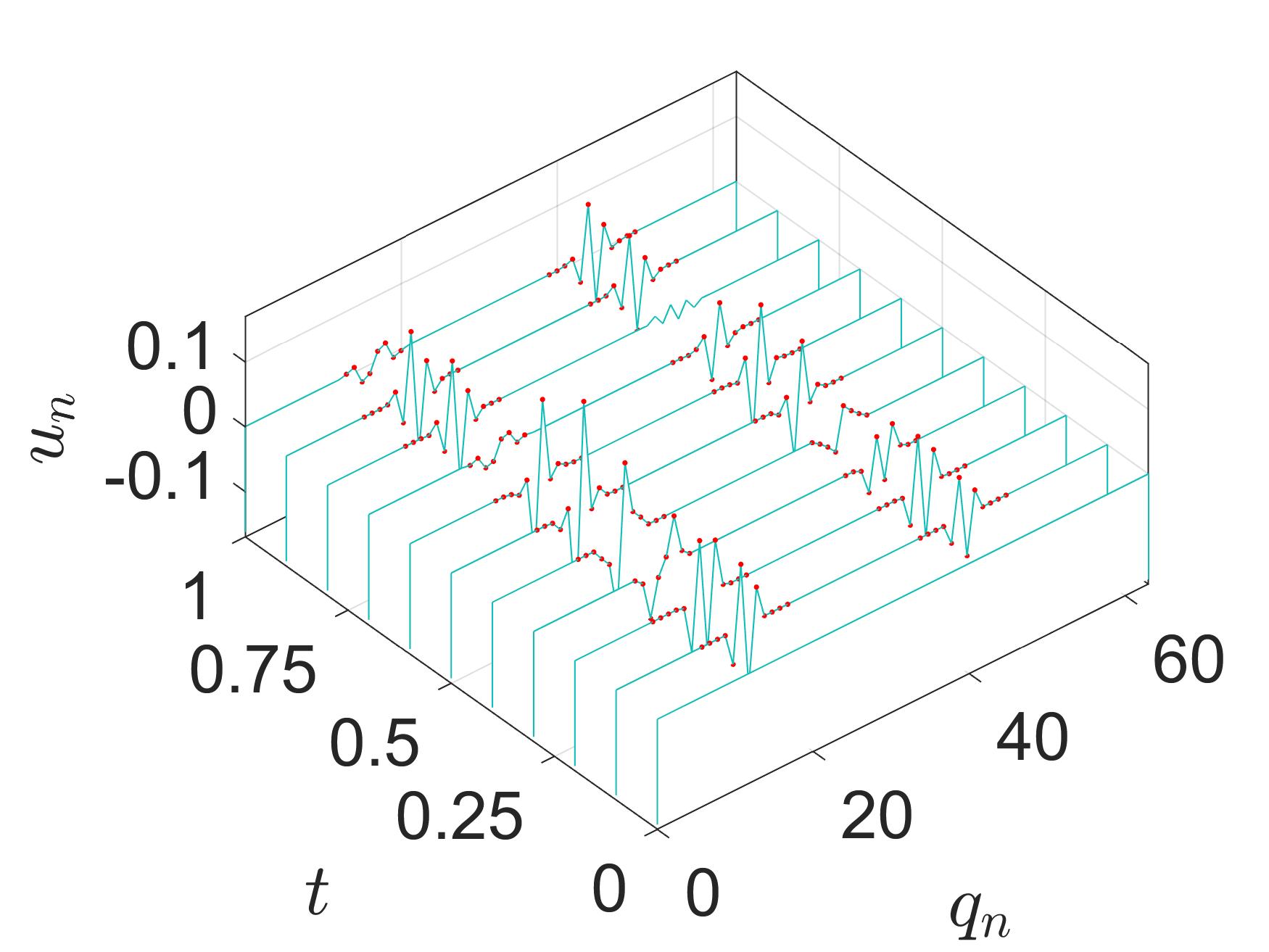}}
\caption{Detection of numerically simulated two stationary breathers excited by momenta pattern \eqref{eq:patern} with $\gamma=0.55$ and $\gamma=0.35$ on the left and right, respectively. Detected localization region particles are identified by red dots. (a) linear SVC with PCA trained on the dataset $X_{u,p,E}$. (b) linear SVC with LLE trained on the dataset $X_{u,p,E}$. (c) nonlinear SVC with PCA trained on the dataset $X_{u}$. (d) nonlinear SVC with LLE trained on the dataset $X_{u}$.}\label{fig:ApplA}
\end{figure}

For this experiment we used four classifiers. In Figure \ref{fig:ApplA_1} we demonstrate results of linear SVC with PCA, while in Figure \ref{fig:ApplA_2} we consider linear SVC with LLE. Both classifiers are trained on the dataset $X_{u,p,E}$ and are able to capture regions of localization very well. As previously suggested, see Figure \ref{fig:dataoffA}, on average SVC classifier with PCA produces wider detection regions, as can be seen from number of colored particles.

In comparison, nonlinear classifiers trained on the dataset $X_{u}$ containing only the information of particle displacements, see Figures \ref{fig:ApplA_3} and \ref{fig:ApplA_4}, are not always able to capture small displacement ILMs. Similar issues (not shown) were observed applying nonlinear classifiers trained on the dataset $X_{u,p}$. Despite these shortcomings, the classifiers can still be applied if additional information of the energy \eqref{eq:En} is unknown.

To further demonstrate methods' robustness and future research directions we consider two additional examples. In the first example we consider a stationary discrete breather in noisy background by initially randomly exciting all particles and their momenta in addition to the pattern \eqref{eq:patern} with $\gamma=0.5$. In this example we performed calculations until time $T_{end}=10$ and used linear SVC with LLE trained on the dataset $X_{u,p,E}$, see Figure \ref{fig:ApplC}. Figure \ref{fig:ApplC} illustrates very well that the method was able to detect the regions of localization despite the presence of large amount of phonon waves. These results suggest that the classification algorithms can be applied to thermostatic dynamics, i.e., during experiments at fixed temperature.

\begin{figure}[t]
\centering 
\subfigure[]{\label{fig:ApplC}
\includegraphics[trim=1.4cm 0.5cm 2cm 3.5cm,clip=true,width=0.48\textwidth]{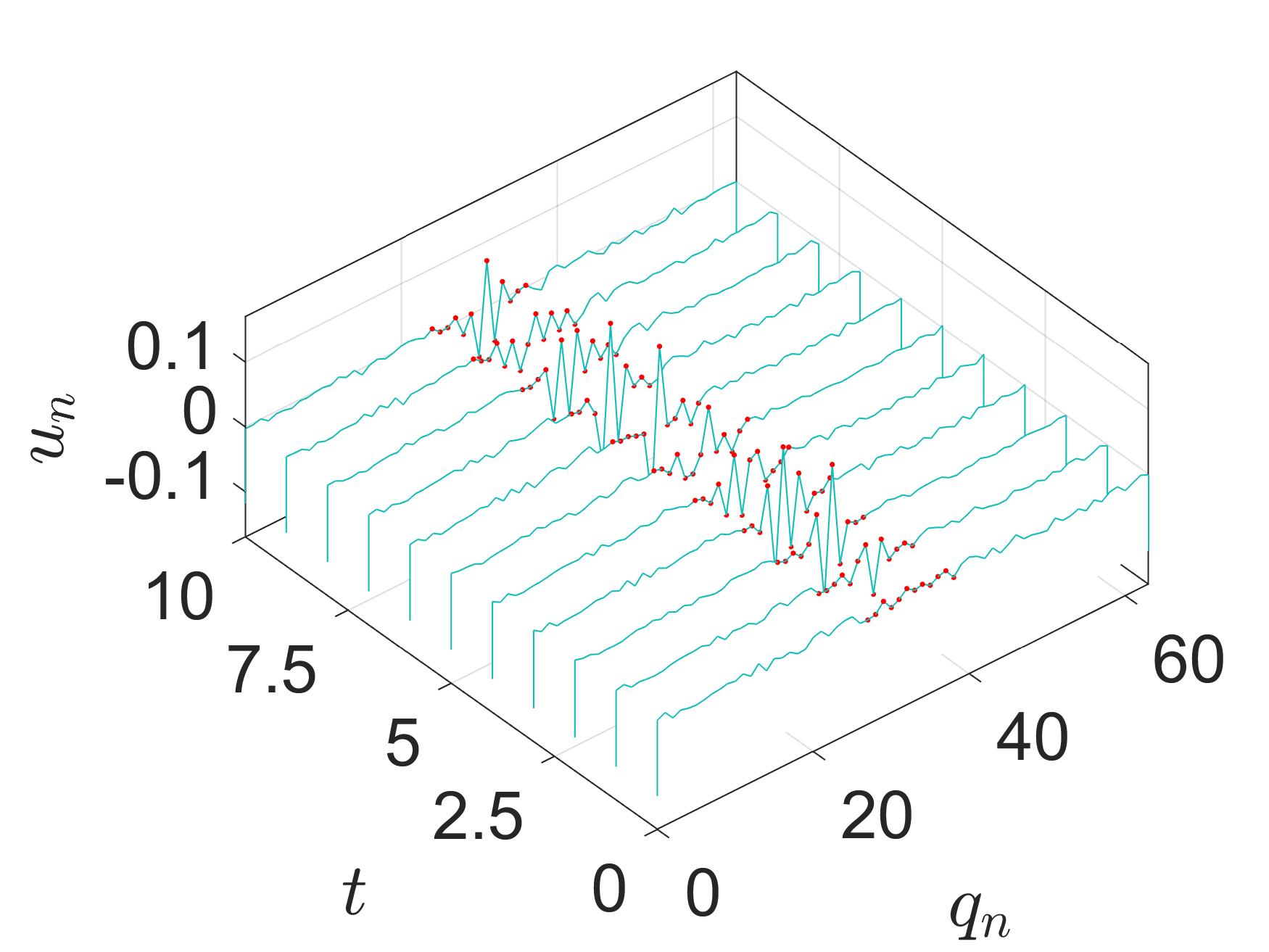}}
\subfigure[]{\label{fig:ApplB}
\includegraphics[trim=1.4cm 0.5cm 2cm 3.5cm,clip=true,width=0.48\textwidth]{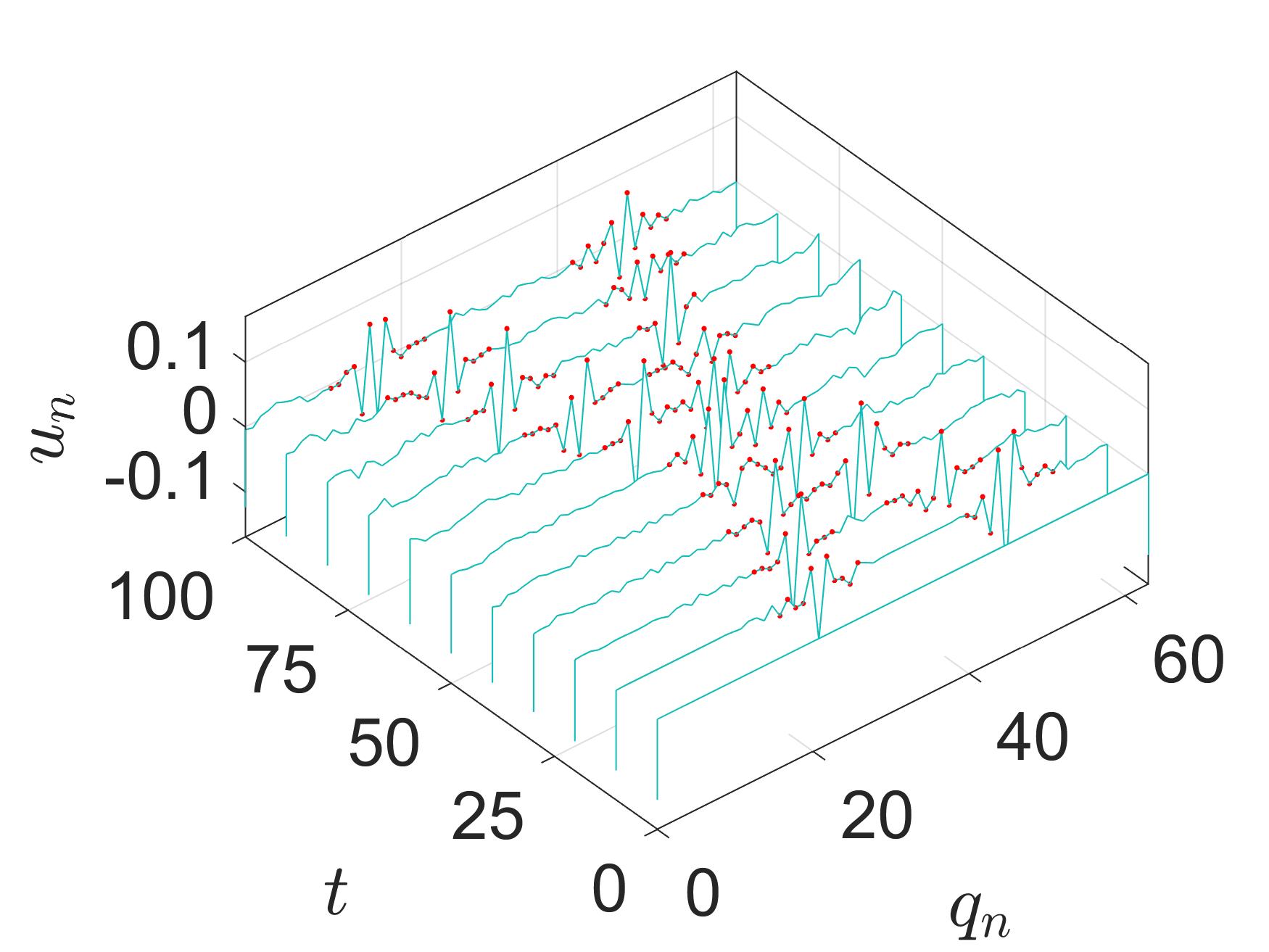}}
\caption{Detection of localization regions in numerical simulations with linear SVC and LLE trained on the dataset $X_{u,p,E}$. Particles of detected regions are identified by red dots. (a) simulation of stationary discrete breather ($\gamma=0.5$) in noisy background. (b) simulation of two mobile breather collision initiated by momenta pattern \eqref{eq:pattern2} with $\gamma=0.45$ (left) and $\gamma=-0.65$ (right).}\label{fig:ApplBC}
\end{figure}

In the second example we consider a simulation of two mobile breather collision initiated by three neighboring particle momenta with pattern:
\begin{equation}\label{eq:pattern2}
\B{p}_0 = \gamma \, (-1, 2, -1)^T.
\end{equation}
The mobile breather propagating to right was excited with $\gamma=0.45$ while the mobile breather propagating to left was excited with $\gamma=-0.65$. Integration in time was performed until $T_{end}=100$. Around time $t=50$ both breathers collide and pass through or reflect from each other and continue to propagate, see Figure \ref{fig:ApplB}. Although the classifier was trained on data obtained from stationary discrete breathers, we are still able to detect the localization regions before, during and after the collision. As in Figure \ref{fig:ApplC} we used linear SVC with LLE trained on the dataset $X_{u,p,E}$. These satisfactory results show that the methodology can be used beyond just detection of stationary DBs.

\section{Discussion and conclusions}\label{sec:Conclusions}

The results of Section \ref{sec:Applications} demonstrate that classification algorithms, such as SVC, either linear or nonlinear, can be used to detect localization regions in numerical simulations of ILMs. The strength of the developed approach lies in the fact that localization regions can be detected from locally sampled data at a given time. In future research we will look for the method which makes predictions with the most optimal width of the localization region and incorporates spectral properties of the lattice waves. For example, in this work we only considered data sampled from eight neighboring particles, other options could also be investigated.

Numerical examples demonstrated that our classifiers are also able to detect mobile DBs but lack the ability to differentiate between mobile and stationary solutions. To address this we will explore multilabel and multioutput classification algorithms in future work. In addition, we plan to extend our work to higher dimensional crystal lattice models and include kink solutions.

For efficiency and visualization purposes with high success we applied two dimensionality reduction methods before training SVC. The best results were obtained with LLE on all datasets containing particle energy data, especially on the dataset $X_{u,p,E}$, such that we could use linear SVC and find more optimal localization regions in numerical experiments. In future work we plan to explore, in particular, more other manifold learning techniques.  

To demonstrate versatility of our approach we considered different datasets for training. We found that the best results are obtained if in addition to particle displacement values particle momenta and energy density values are also added to the dataset. Nevertheless, classifiers can be built on all the datasets and overall good performance can be achieved.

\section*{Acknowledgements}
J.~Baj\={a}rs acknowledges support from PostDocLatvia grant No.1.1.1.2/VIAA/4/20/617.

\bibliographystyle{unsrt}  
\bibliography{mybibfile}  

\end{document}